\documentclass[12pt]{article}

\usepackage{bbm}
\usepackage{bm}
\usepackage{amsmath}
\usepackage{amsfonts}

\begin{document}
\pagestyle{plain}

\makeatletter
\@addtoreset{equation}{section}
\makeatother
\renewcommand{\theequation}{\thesection.\arabic{equation}}
\pagestyle{empty}
\rightline{CPHT-RR152.1007}
\rightline{LPT-ORSAY 07-95}
\rightline{ROM2F/2007/14}
\begin{center}
\LARGE{String instantons, fluxes and moduli stabilization
\\[10mm]}
\large{ P.G. C\'amara${}^a$,
E. Dudas${}^{a,b}$, T. Maillard${}^a$
and G. Pradisi${}^c$ \\[6mm]}
\small{
${}^a$Centre de Physique Th\'eorique\footnote{Unit\'e mixte du CNRS, UMR 7644.}, Ecole Polytechnique, F-91128 Palaiseau, France. \\
${}^b$LPT\footnote{Unit\'e mixte du CNRS, UMR 8627.}, Bat. 210, Univ. de Paris-Sud, F-91405 Orsay, France.\\
${}^c$Dipartimento di Fisica, Univ. di Roma ``Tor Vergata'' and INFN - Sez. Roma II,\\[-0.3em]
Via della Ricerca Scientifica 1, 00133 Roma, Italy
\\[1cm]}
\small{\bf Abstract} \\[0.5cm]
\end{center}
{\small
We analyze a class of dual pairs of heterotic and type I models based on freely-acting
$\mathbb{Z}_2 \times \mathbb{Z}_2$ orbifolds in four dimensions.  Using the adiabatic argument, it is possible to calculate non-perturbative contributions to the gauge coupling threshold corrections on the type I side by
exploiting perturbative calculations on the heterotic side, without the drawbacks due to twisted moduli.  The instanton effects can then be
combined with closed-string fluxes to stabilize most of the moduli fields of the internal manifold, and also the dilaton, in a racetrack realization of the type I model.}

\newpage
\setcounter{page}{1}
\pagestyle{plain}
\renewcommand{\thefootnote}{\arabic{footnote}}
\setcounter{footnote}{0}
\tableofcontents

\section{Introduction and Conclusions}

In recent years new ways to compute non-perturbative effects in
string theory were developed, based on Euclidean $p$-branes
($Ep$-branes) wrapping various cycles of the internal manifold of
string compactifications \cite{witten,ganor,billo,bcw,iu,kachru,other}.
Some of the instanton effects have an interpretation in terms of
gauge theory instantons, whereas others are stringy instanton
effects whose gauge theory counterpart is still under investigation
(For recent reviews on instanton effects in field and string theory,
see e.g. \cite{instantonreviews}). Whereas the former effects are
responsible for the generation of non-perturbative superpotentials
via gauge theory strong IR dynamics \cite{tvy} and of moduli
potentials satisfying various gauge invariance constraints
\cite{bd}, the latter could be responsible for
generating Majorana neutrino masses or the $\mu$-term in MSSM
\cite{bcw,iu}, as well as for inducing other interesting effects at
low energy \cite{other}.

  The purpose of the present paper is to present a class of examples based on freely-acting
$\mathbb{Z}_2 \times \mathbb{Z}_2$ orbifold models, that adds two new ingredients to the discussion,
trying to go deeper into the
non-perturbative effects analysis. The first new ingredient is the heterotic-type I
duality \cite{witten2}, which exchanges perturbative and
non-perturbative regimes. As is well known \cite{vw}, it is possible
to construct freely-acting dual pairs with ${\cal N}=1$
supersymmetry in four dimensions which preserve the S-duality structure. As we show
explicitly here, the dual pairs can have a rich non-perturbative
dynamics exhibiting both types of effects mentioned above. The heterotic-type I
duality allows, for example, to obtain the exact $E1$
instantonic summations on the type I side for the
non-perturbative corrections to the gauge couplings using the computation of
perturbative
threshold corrections on the heterotic side\footnote{See \cite{bfkov} for
earlier work on instanton effects and heterotic-type I duality.}.
Second, non-perturbative effects also play a potentially important role
in addressing the moduli field stabilization issue.  Closed string fluxes
were invoked in recent years in the framework of type IIB and type IIA
string compactifications,
following the initial proposal of \cite{gkp} to try to stabilize all moduli
fields, including the dilaton.
The combination of closed string fluxes and freely-acting orbifold actions has
the obvious advantage of avoiding to deal with twisted-sector moduli
fields, absent in our construction.  We show that, besides the Ramond-Ramond (RR) three-form fluxes, also metric
fluxes can be turned on in our freely-acting type I models,
requiring new quantization conditions and the twisting of the cohomology
of the internal manifold.  The low-energy effective description is equivalent to the original
one, with the addition of a non-trivial superpotential.  Moreover, our
string constructions allow naturally racetrack models with
dilaton stabilization \cite{krasnikov}.  We show how they can be combined with
closed string fluxes and stringy instanton effects in order to
stabilize most of the moduli fields of the internal manifold.

The plan of the paper is as follows.  In Section 2 we discuss the geometric
framework of the freely acting $\mathbb{Z}_2 \times \mathbb{Z}_2$ orbifolds.  In Section 3 we display the explicit type I descendants obtained by quotienting the orbifold with the geometric
world-sheet parity operator.  Besides some variations of the simplest
class with orthogonal gauge groups, we also construct the
corresponding heterotic duals in Section 4.  In Section 5, we report
the calculation of
the threshold corrections to the gauge couplings both for the
heterotic and for the type I models.
The details of the calculations are reported in the Appendices.  In
particular, we verify that the
moduli dependence of the non-perturbative corrections on the type I
side is in agreement with the
conjectured form \cite{germans}.  In Section 6 we analyze the
instanton contributions in the type I framework,
that are combined with closed string fluxes in Section 7 in order to attain the stabilization of
most of the moduli of the compactification manifold.  In particular, in Section 7 we describe
an example in which the
dilaton can be also stabilized, due to a natural racetrack realization of the type I model in combination with closed
metric and RR three-form fluxes.

\section{The freely-acting orbifold and its moduli}

From the point of view of the target space, we take a $T^6$ torus
($y^i = y^i +1$) with vielbein vectors $e^i=e^i{}_{\mu}dy^{\mu}$ and
metric given by
\begin{equation}
ds^2=\sum_ie^i{}_{\mu}e^{i\mu}\ .
\end{equation}

This has to be $SL(6,\mathbb{Z})$ invariant. Therefore, performing a general rotation of the lattice
vectors one may write a basis as follows\footnote{We use the notation $y^i$, $i=1\ldots 6$, to denote
the internal compact dimensions and $x^i$, $i=0\ldots 3$, for the non-compact space-time dimensions.}
\begin{align}
e^6&=R^6dy^6 \ , \label{e6} \\
e^5&=R^5(dy^5+a^5{}_6dy^6) \ , \\
e^4&=R^4(dy^4+a^4{}_5dy^5+a^4{}_6dy^6) \ , \\
e^3&=R^3(dy^3+a^3{}_4dy^4+a^3{}_5dy^5+a^3{}_6dy^6) \ , \\
e^2&=R^2(dy^2+a^2{}_3dy^3+a^2{}_4dy^4+a^2{}_5dy^5+a^2{}_6dy^6) \ , \\
e^1&=R^1(dy^1+a^1{}_2dy^2+a^1{}_3dy^3+a^1{}_4dy^4+a^1{}_5dy^5+a^1{}_6dy^6) \ . \label{e1}
\end{align}

Modding by the orbifold action will break the $SL(6,\mathbb{Z})$
symmetry to a smaller subgroup. We define the generators $\{g,f,h\}$
of the $\mathbb{Z}_2 \times \mathbb{Z}_2$ freely-acting orbifold as,
\begin{align}
&(y^1,y^2,y^3,y^4,y^5,y^6) \xrightarrow{g} (y^1+1/2,y^2,-y^3,-y^4,-y^5+1/2,-y^6) \ , \label{gen1}\\
&(y^1,y^2,y^3,y^4,y^5,y^6) \xrightarrow{f} (-y^1+1/2,-y^2,y^3+1/2,y^4,-y^5,-y^6) \ , \\
&(y^1,y^2,y^3,y^4,y^5,y^6) \xrightarrow{h}
(-y^1,-y^2,-y^3+1/2,-y^4,y^5+1/2,y^6) \ . \label{gen3}
\end{align}
Notice that these orbifold operations have no fixed points due to the
shifts, hence they act freely (see e.g. \cite{kk}). Moreover, for objects
localized in the internal space, as will be the case for the $E1_i$
instantons to be discussed in Section 6, orbifold operations will
generate inevitably instanton images.  This has
non-trivial consequences on the instanton spectra, as we shall see later on.

In order for the lattice vectors (\ref{e6}) - (\ref{e1}) to transform covariantly with
respect to the orbifold action, it is required that
\begin{equation}
a^4{}_5=a^4{}_6=a^3{}_5=a^3{}_6=a^2{}_3=a^2{}_4=a^2{}_5=a^2{}_6=a^1{}_3=a^1{}_4=a^1{}_5=a^1{}_6=0\ .
\end{equation}
A basis of holomorphic vectors can thus be introduced in the form
\begin{align}
&z^1=e^1+ie^2=R_1(dy^1+iU_1dy^2)\ , \label{ez1} \\
&z^2=e^3+ie^4=R_3(dy^3+iU_2dy^4)\ , \\
&z^3=e^5+ie^6=R_5(dy^5+iU_3dy^6)\ , \label{ez3}
\end{align}
where we have defined
\begin{equation}
U_1=\frac{R^2}{R^1}-ia^1{}_2 \quad , \quad U_2=\frac{R^4}{R^3}-ia^3{}_4 \quad , \quad U_3=\frac{R^6}{R^5}-ia^5{}_6\ .
\end{equation}

Hence, the moduli space of the untwisted sector matches precisely  the one of an
ordinary $\mathbb{Z}_2\times \mathbb{Z}_2$, given by the three complex
structure moduli, $U_i$, together with the three K\"ahler moduli, $T_i$,
which result from the expansion of the
complexified K\"ahler 2-form in a cohomology basis of even 2-forms,
\begin{equation}
J_c = e^{-\phi} J + i C_2 =
T_1 dy^1\wedge dy^2 + T_2 dy^3\wedge dy^4 + T_3 dy^5\wedge dy^6 \ .
\end{equation}
Making use of (\ref{ez1}) - (\ref{ez3}), the real parts of
the K\"ahler moduli can be seen to be
\begin{equation}
\textrm{Re }T_1=e^{-\phi}R_1R_2\quad , \quad \textrm{Re }T_2=e^{-\phi}R_3R_4\quad , \quad \textrm{Re }T_3=e^{-\phi}R_5R_6\ .
\end{equation}
The effective theory contains also, as usual, the universal
axion-dilaton modulus
\begin{equation}
S \ = \  e^{-\phi} \prod_{i=1}^6 R_i \ + \ i \ c \ , \label{moduli1}
\end{equation}
where $c$ is the universal axion.
On the other hand, since there are no fixed points in the orbifold
action, we expect the twisted sector to be trivial. We shall see in next section, from the exchange
of massless modes in the vacuum amplitudes, that this is indeed the
case. The internal space of the orbifold is therefore completely
smooth and can be interpreted as a Calabi-Yau space with Hodge numbers
$(h_{11},h_{21}) = (3,3)$. The corresponding type IIB string theory on
this orbifold space has the standard left-right worldsheet involution
$\Omega_P$ as a symmetry, which we use, following \cite{partialbreak,aads2},
in order to construct type I freely-acting orbifolds.
\section{Type I  models~: vacuum energy and spectra}

\subsection{Type I with orthogonal gauge groups}
\label{typeImodels}

We briefly summarize here some of the results of
\cite{partialbreak}.   Following the original notation, the $\mathbb{Z}_2
\times \mathbb{Z}_2$ orbifold generators of eqs. (\ref{gen1}) - (\ref{gen3}) can be written as
\begin{equation}
g = (P_1, -1, -P_3) \quad, \quad f = (-P_1, P_2, -1) \quad , \quad h =
(-1, -P_2, P_3) \ , \label{ort1}
\end{equation}
where $P_i$ represents the momentum shift along the real direction $y^{2 i -1}$ of the i-th torus.
We consider the type I models obtained by gauging the type IIB string with $\Omega_P$,
the standard worldsheet orientifold involution.
The spectrum can be read from the one-loop amplitudes \cite{rome}. In particular,
the torus partition function is\footnote{ There is an overall normalization that is explicitly written in Appendix \ref{normaliz}.
For other conventions concerning orientifolds, see e.g. the reviews \cite{as}. }
\begin{multline}
T=\int_{\cal F} \frac{d^2 \tau}{\tau_2^3 |\eta|^4}\frac{1}{4}\left[|\tau_{oo}+\tau_{og}+\tau_{oh}+\tau_{of}|^2\Lambda_1\Lambda_2\Lambda_3+
|\tau_{oo}+\tau_{og}-\tau_{oh}-\tau_{of}|^2(-1)^{m_1}\Lambda_1\left|\frac{4\eta^2}{\vartheta_2^2}\right|^2\right. \\
\left.+\ |\tau_{oo}-\tau_{og}+\tau_{oh}-\tau_{of}|^2(-1)^{m_3}\Lambda_3\left|\frac{4\eta^2}{\vartheta_2^2}\right|^2+
|\tau_{oo}-\tau_{og}-\tau_{oh}+\tau_{of}|^2(-1)^{m_2}\Lambda_2\left|\frac{4\eta^2}{\vartheta_2^2}\right|^2+\right. \\
\left. +\ |\tau_{go}+\tau_{gg}+\tau_{gh}+\tau_{gf}|^2\Lambda_1^{n_1+\frac{1}{2}}\left|\frac{4\eta^2}{\vartheta_4^2}\right|^2+
|\tau_{go}+\tau_{gg}-\tau_{gh}-\tau_{gf}|^2(-1)^{m_1}\Lambda_1^{n_1+\frac{1}{2}}\left|\frac{4\eta^2}{\vartheta_3^2}\right|^2+\right. \\
\left. +\ |\tau_{ho}+\tau_{hg}+\tau_{hh}+\tau_{hf}|^2\Lambda_3^{n_3+\frac{1}{2}}\left|\frac{4\eta^2}{\vartheta_4^2}\right|^2+
|\tau_{ho}-\tau_{hg}+\tau_{hh}-\tau_{hf}|^2(-1)^{m_3}\Lambda_3^{n_3+\frac{1}{2}}\left|\frac{4\eta^2}{\vartheta_3^2}\right|^2+\right. \\
\left. +\ |\tau_{fo}+\tau_{fg}+\tau_{fh}+\tau_{ff}|^2\Lambda_2^{n_2+\frac{1}{2}}\left|\frac{4\eta^2}{\vartheta_4^2}\right|^2+
|\tau_{fo}-\tau_{fg}-\tau_{fh}+\tau_{ff}|^2(-1)^{m_2}\Lambda_2^{n_2+\frac{1}{2}}\left|\frac{4\eta^2}{\vartheta_3^2}\right|^2\right]\ ,
\label{tso32}
\end{multline}
while the Klein bottle, annulus and
M\"obius strip amplitudes read in the direct (loop) channel respectively as
\begin{align}
K&=\int_0^{\infty} \frac{dt}{t^3\eta^2}\frac{1}{8}(\tau_{oo}+\tau_{og}+\tau_{oh}+\tau_{of})\nonumber \\
&\{P_1P_2P_3\ +\ (-1)^{m_1}P_1W_2W_3\ +\ W_1(-1)^{m_2}P_2W_3\ +\ W_1W_2(-1)^{m_3}P_3\} \ , \label{kso32}
\end{align}
\begin{align}
A&=\int_0^{\infty} \frac{dt}{t^3\eta^2}\frac{1}{8}\nonumber \\
&\{ I_N^2(\tau_{oo}+\tau_{og}+\tau_{oh}+\tau_{of})P_1P_2P_3\ +\
g_N^2(\tau_{oo}+\tau_{og}-\tau_{oh}-\tau_{of})(-1)^{m_1}P_1\frac{4\eta^2}{\vartheta_2^2}\ + \label{aso32}\\
&h_N^2(\tau_{oo}-\tau_{og}+\tau_{oh}-\tau_{of})(-1)^{m_3}P_3\frac{4\eta^2}{\vartheta_2^2}\ +\
f_N^2(\tau_{oo}-\tau_{og}-\tau_{oh}+\tau_{of})(-1)^{m_2}P_2\frac{4\eta^2}{\vartheta_2^2}\}
\ , \nonumber
\end{align}
\begin{align}
M&=-\int_0^{\infty} \frac{dt}{t^3\eta^2}\frac{I_N}{8}\nonumber \\
&\{ (\hat{\tau}_{oo}+\hat{\tau}_{og}+\hat{\tau}_{oh}+\hat{\tau}_{of})P_1P_2P_3\ +\
(\hat{\tau}_{oo}+\hat{\tau}_{og}-\hat{\tau}_{oh}-\hat{\tau}_{of})(-1)^{m_1}P_1\frac{4\hat{\eta}^2}{\hat{\vartheta}_2^2}\ +\nonumber \\
&(\hat{\tau}_{oo}-\hat{\tau}_{og}+\hat{\tau}_{oh}-\hat{\tau}_{of})(-1)^{m_3}P_3\frac{4\hat{\eta}^2}{\hat{\vartheta}_2^2}\ +\
(\hat{\tau}_{oo}-\hat{\tau}_{og}-\hat{\tau}_{oh}\ +\ \hat{\tau}_{of})(-1)^{m_2}P_2\frac{4\hat{\eta}^2}{\hat{\vartheta}_2^2}\} \ . \label{mso32}
\end{align}
Some comments on the notation are to be made. In the torus
amplitude, ${\cal F}$ is the fundamental domain and the $\Lambda_i$
are the lattice sums for the three compact tori, whereas the
shorthand notation $(-1)^{m_i}\Lambda_i^{n_i+1/2}$ indicates a sum
with the insertion of $(-1)^{m_i}$ along the momentum in $y^{2i-1}$,
with the corresponding winding number shifted by $1/2$. $P_i$ and
$W_i$ in (\ref{kso32}) - (\ref{mso32}) are respectively the momentum
and winding sums for the three two-dimensional tori. More
concretely, using for the geometric moduli the conventions of the
previous section, one has\footnote{In what follows we set the string tension
$\alpha'=1/2$.}
\begin{align}
P_i&\equiv \sum_{m,m'}\textrm{exp}\left[-\frac{\pi t}{(\textrm{Re }T_i)(\textrm{Re }U_i)}|m'-iU_im|^2\right] \ , \\
(-1)^{m_i}P_i&\equiv \sum_{m,m'}(-1)^m\textrm{exp}\left[-\frac{\pi t}{(\textrm{Re }T_i)(\textrm{Re }U_i)}|m'-iU_im|^2\right]\ . \label{mp}
\end{align}
Moreover, in (\ref{mso32}) hatted modular functions define a
correct basis under the P transformation extracting a suitable
overall phase~\cite{rome}. Indeed, the moduli of the
double-covering tori are $\tau = (i t/2 + 1/2)$ for the
M\"obius-strip amplitude, $\tau = 2 i t$ for the
Klein-bottle amplitude and $\tau = i t/2$ for the annulus
amplitude. In Appendix \ref{apcharacters} we give the definition
of the characters used in eqs. (\ref{kso32}) - (\ref{mso32}) in terms of $[SO(2)]^4$ characters.

It is worth to analyze the effects of the freely-acting
operation on the geometry of the models.
In general, $\mathbb{Z}_2 \times \mathbb{Z}_2$
orientifolds contain $O9$-planes and three sets of $O5_i$-planes defined as
the fixed tori of the operations $\Omega_P \circ g$, $\Omega_P \circ f$, $\Omega_P \circ h$, each wrapping
one of the three internal tori $T^i$. In our freely-acting
orbifold case, the overall $O5_i$-plane charges are zero and the
$O5_i$-planes couple only to massive (odd-windings) states. A
geometric picture of this fact can be obtained T-dualizing the two directions the $O5_i$ planes wrap, so that they become
$O3_i$-planes. In this way, the freely acting operation replaces
the $O3_{i,-}$ planes by ($O3_{i,+}$- $O3_{i,-}$) pairs,
separated by half the lattice spacing in the coordinate affected by
the free action. Since there are no global background charges  from $O5_{i}$-planes, the
model contains only background $D9$ branes.
Finally, the
Chan-Paton $D9$ charges are defined as,
\begin{align}
I_N& = n_o + n_g + n_h + n_f \ , &  g_N& = n_o + n_g - n_h - n_f \ , \nonumber  \\
h_N& = n_o - n_g + n_h - n_f \ , &  f_N& = n_o - n_g - n_h + n_f \ , \label{cp1}
\end{align}
with $I_N=32$ fixed by the tadpole cancellation condition. The
massless spectrum has ${\cal N}=1$ supersymmetry. The gauge group is
$SO(n_o) \otimes SO(n_g) \otimes SO(n_h) \otimes SO(n_f)$,
with chiral multiplets in the bifundamental representations
\begin{eqnarray}
&& {\bf (n_o, n_g, 1,1 )}\ +\ {\bf (n_o, 1, n_f, 1 )}\ +\ {\bf (n_o, 1, 1,
  n_h )} \ + \nonumber \\
&& + \ {\bf (1, n_g, n_f, 1 )}\ +\ {\bf (1,n_g, 1,n_h )}\ +\ {\bf (1, 1,
  n_f, n_h )} \ . \label{ort2}
\end{eqnarray}
The existence of four different Chan-Paton charges can be traced  to the
various consistent actions of the orbifold group on the Chan-Paton space or, alternatively, to the number of independent sectors of the chiral Conformal Field Theory.
It can be useful for the reader to make a connection with the
alternative notation of \cite{gp}. The original Chan-Paton charges can be
grouped into a $32 \times 32$ matrix $\lambda$. In this Chan-Paton matrix
space, the three orbifold operations $g,f$ and $h$ act via matrices
$\gamma_g, \gamma_f, \gamma_h$ which, correspondingly to (\ref{cp1}), are given by
\begin{align}
\gamma_g \ = \ (I_{n_o}, I_{n_g}, - I_{n_f}, -I_{n_h}) \ , \ \nonumber \\
\gamma_f \ = \ (I_{n_o}, - I_{n_g}, I_{n_f}, -I_{n_h}) \ , \ \label{cp3} \\
\gamma_h \ = \ (I_{n_o}, - I_{n_g}, - I_{n_f}, I_{n_h}) \ , \ \nonumber
\end{align}
where $I_{n_o}$ denote the identity matrix in the $n_o \times n_o$
block diagonal Chan-Paton matrix, and the same for the other
multiplicities $n_i$. For $n_g=n_h=n_f=0$ one recovers a pure
$SO(32)$ SYM with no  extra multiplets, a theory where gaugino
condensation is expected to arise. Finally, let us notice that
even if perturbatively $n_o,n_g,n_f,n_h$ can be arbitrary positive
integers subject only to the tadpole condition
$n_o+n_g+n_f+n_h=32$, non-perturbative consistency asks all of
them to be even integers.

\subsection{Type I racetrack model}

In a variation of the previous $SO(32)$ model, we may add a
discrete deformation along one of the unshifted directions, similar to a Wilson line $A_2 =
(e^{2 \pi i {\bf a}})$ along $y^2$, with ${\bf a} = ({\bf 0}_p, {\bf 1/2}_{32-p})$ and breaking $SO(32)
\rightarrow SO(p) \otimes SO(32-p)$. The annulus
and M\"obius amplitudes, (\ref{aso32}) and (\ref{mso32}), get correspondingly modified to the
following expressions:
\begin{align}
&A=\int_0^{\infty} \frac{dt}{t^3\eta^4}\frac{1}{8}\{[(p^2+q^2)P_{m_1'}
 + 2pqP_{m_1'+\frac{1}{2}}]P_{m_1}P_2P_3(\tau_{oo}+\tau_{og}+\tau_{oh}+\tau_{of})\ +\nonumber \\
&+(p^2+q^2)[(-1)^{m_2}P_2(\tau_{oo}-\tau_{og}-\tau_{oh}+\tau_{of}) +
(-1)^{m_3}P_3(\tau_{oo}-\tau_{og}+\tau_{oh}-\tau_{of})]\frac{4\eta^2}{\vartheta_2^2} \nonumber \\
&\quad +(-1)^{m_1}[(p^2+q^2)P_{m_1'}+
2pqP_{m_1'+\frac{1}{2}}]P_{m_1}(\tau_{oo}+\tau_{og}-\tau_{oh}-\tau_{of})
\frac{4\eta^2}{\vartheta_2^2}\} \ ,
\label{raa}
\end{align}
\begin{align}
&M=-\frac{p+q}{8}\{P_1P_2P_3({\hat \tau_{oo}}+{\hat \tau_{og}}+{\hat \tau_{oh}}+\tau_{of})\ +\ (-1)^{m_1}P_1
({\hat \tau_{oo}}+{\hat \tau_{og}}-{\hat \tau_{oh}}-{\hat \tau_{of}})\frac{4\eta^2}{\vartheta_2^2}\nonumber \\
&+\ (-1)^{m_2}P_2({\hat \tau_{oo}}-{\hat \tau_{og}}-{\hat \tau_{oh}}+{\hat \tau_{of}})
\frac{4\eta^2}{\vartheta_2^2}\ +\
(-1)^{m_3}P_3({\hat \tau_{oo}}-{\hat \tau_{og}}+{\hat \tau_{oh}}-{\hat \tau_{of}})\frac{4\eta^2}{\vartheta_2^2}\} \ . \label{ram}
\end{align}
As mentioned, $I_N=p+q=32$,
\begin{align}
P_{m_1'}P_{m_1}& \equiv P_1 \ , \\
{\rm and} \nonumber \\
P_{m_1'+\frac{1}{2}}P_{m_1}& \equiv \sum_{m,m'}\textrm{exp}\left[-\frac{\pi t}{(\textrm{Re }T_1)(\textrm{Re }U_1)}|m'-iU_1m+1/2|^2\right] \ .
\end{align}
Hence, the resulting $SO(p) \otimes SO(32-p)$ gauge group is accompanied by a
pure ${\cal N}=1$ SYM theory
on both factors, leading to a racetrack scenario with two gaugino condensates.
Indeed, in the four-dimensional effective supergravity Lagrangian, the tree-level gauge
kinetic functions on the two stacks of $D9$ branes are equal,
\begin{equation}
f_{SO(p)} \ = \ f_{SO(q)} \ = \ S \ , \label{race1}
\end{equation}
where $S$ is the universal dilaton-axion chiral multiplet. Gaugino
condensation on both stacks then generates the non-perturbative
superpotential
\begin{equation}
W_{np} \ = \ A_p^{(k)} e^{- {a_p S }} + A_q^{(l)} e^{- {a_q S }} \ ,
\label{race2}
\end{equation}
where $A_p^{(k)} = (p-2) \ \exp (2 \pi i k/(p-2))$ and  $A_q^{(l)} =
(q-2) \ \exp (2 \pi i l/(q-2))$, with $k =  1 \ldots p-2$ and $l = 1
\ldots q-2$, provide the requested different phases of the SYM vacua
\cite{vy}. Moreover, $a_p = 2/(p-2)$ ($a_q = 2/(q-2)$) is related to
the one-loop beta function of the $SO(p)$ ($SO(q)$) SYM gauge
factor. In addition to the massless states, the model contains
massive states, in particular a massive vector multiplet in the ${\bf (p, q)}$
bifundamental representation, with a lowest mass of the order of
the compactification scale $M_c \sim 1/R$. Since the
four-dimensional effective theory is valid anyway below $M_c$, these
states are heavy and their effects on the low-energy physics can be
encoded in threshold effects which we shall compute later on.

An interesting question is the geometrical interpretation of the
present model\footnote{E.D. is grateful to C.
Angelantonj and M. Bianchi for illuminating discussions on this
and the other string models presented in the present paper.}.
The natural interpretation is in terms of a Wilson
line breaking of the $SO(p) \otimes SO(32-p)$ model.  The absence of scalars describing
positions of the branes corresponding to each $SO$ factor indicates that the
corresponding branes are fractional and, as such, cannot move outside the fixed
points.  However, by giving vev's to the scalars in bifundamentals, one
converts fractional branes into regular branes.  The resulting gauge group is $SO(2P) \otimes SO(16-P)$, where the
first factor comes from the branes sitting at the fixed point, while the second factor describe brane pairs
in the bulk having scalars in the symmetric representation corresponding to their positions.
Moving the bulk branes to another fixed-point, one
gets, as usual, an enhancement of the gauge group to $SO(2P) \otimes SO(32-2P)$.


 \subsection{Type I with unitary groups}

It is interesting to analyze the non-perturbative dynamics of the
gauge theory on the $D9$ branes in the case of an orbifold action
on the Chan-Paton space that produces unitary gauge groups. This
can be done in a very simple way by choosing a different
Chan-Paton assignment compared to (\ref{cp1}). Consider the same
cylinder amplitude (\ref{aso32}) equipped with the following
parametrization of the Chan-Paton charges:
\begin{eqnarray}
&& I_N = n + {\bar n} + m + {\bar m} \quad , \quad g_N = n + {\bar n}
- m - {\bar m} \ , \nonumber \\
&&  f_N = i ( n - {\bar n} + m - {\bar m}) \quad , \quad h_N = i ( n - {\bar n}
- m + {\bar m}) \ . \label{uni1}
\end{eqnarray}
The M\"obius amplitude has to be changed for consistency into
\begin{multline}
M=-\int_0^{\infty} \frac{dt}{t^3\eta^4}\frac{I_N}{8}
\{ (\hat{\tau}_{oo}+\hat{\tau}_{og}+\hat{\tau}_{oh}+\hat{\tau}_{of})P_1P_2P_3\ +\
(\hat{\tau}_{oo}+\hat{\tau}_{og}-\hat{\tau}_{oh}-\hat{\tau}_{of})(-1)^{m_1}P_1\frac{4\hat{\eta}^2}{\hat{\vartheta}_2^2} \ \\
- (\hat{\tau}_{oo}-\hat{\tau}_{og}+\hat{\tau}_{oh}-\hat{\tau}_{of})(-1)^{m_3}P_3\frac{4\hat{\eta}^2}{\hat{\vartheta}_2^2}\ - \
(\hat{\tau}_{oo}-\hat{\tau}_{og}-\hat{\tau}_{oh}\ +\ \hat{\tau}_{of})(-1)^{m_2}P_2\frac{4\hat{\eta}^2}{\hat{\vartheta}_2^2}\} \ , \label{uni2}
\end{multline}
where the changes of sign in the $D9$-$O5_{2}$ and  $D9$-$O5_{3}$
propagation, needed to enforce the unitary
projection, are interpreted as discrete Wilson lines on
the $D9$ branes in the last two torii \cite{rome}. The massless open string
amplitudes,
\begin{multline}
A_0 + M_0 = ( n {\bar n} + m {\bar m} ) \tau_{oo} + \left[ \frac{n (n-1)}{2}
+\frac{{\bar n} ({\bar n}-1)}{2} + \frac{m (m-1)}{2} +
\frac{{\bar m} ({\bar m}-1)}{2} \right] \tau_{og} \\
+ ( n {\bar m} + {\bar n} m ) \tau_{of} + ( n m + {\bar n} {\bar m}
) \tau_{oh} \ ,  \label{uni3}
\end{multline}
exhibit the spectrum of an ${\cal N}=1$ supersymmetric $U(n) \otimes U(m)$ theory, with $n+m=16$
due to the $(D9/O9)$ RR tadpole
cancellation condition. Matter fields fall into  massless chiral multiplets in the
representations
\begin{multline}
{\bf \left( \frac{n (n-1)}{2} +\frac{{\bar n} ({\bar n}-1)}{2} , 1 \right)}\
+\ {\bf \left( 1, \frac{m (m-1)}{2} +\frac{{\bar m} ({\bar m}-1)}{2} \right)}\
+ \\
+\ {\bf (n, {\bar m})}\ +\ {\bf ({\bar n}, m)}\ +\ {\bf (n,m)}\ +\
{\bf ({\bar n}, {\bar m})} \ . \ \ \ \ \ \ \ \ \ \ \ \label{uni4}
\end{multline}
Notice that the choice $m=0$ with a gauge group $U(16)$, in contrast to the
$SO(32)$ case, is not pure SYM, since it contains massless chiral
multiplets in the (${\bf 120} + {\bf {\overline{120}}}$) representation.

The gauge theory on $D9$ branes is not really supersymmetric QCD with
flavors in the fundamental and antifundamental representation, whose
non-perturbative dynamics is known with great accuracy \cite{tvy}.
One way to get a more interesting example is the following. Moving
$p$ $D9$ branes out of the total $16$ to a different orientifold
fixed point not affected by the shift, one gets a gauge group
$U(n) \otimes U(m) \otimes U(p)$, with $n+m+p=16$. Strings
stretched between the $p$ $D9$ branes and the remaining $n+m$ are
massive, and therefore they disappear from the effective
low-energy gauge theory, whereas the $U(n) \otimes U(m)$ gauge
sector has the massless spectrum displayed in
(\ref{uni4}). Choosing $n=3$ and $m=1$, a gauge group $SU(3)
\otimes U(1)^2$ results, together with a factor $U(12)$ decoupled
from it. Using the fact that the antisymmetric representation of
$SU(3)$ coincides with the antifundamental ${\bar 3}$, one ends up with a SQCD theory with gauge group $SU(3)$ and $N_f=3$ flavors of
quarks-antiquarks. This is the regime $N_c=N_f= N$ described in
\cite{seiberg}, where the composite mesons $M = Q {\bar Q} $ and
baryons (antibaryons) $B = Q_1 \cdots Q_n$  (${\tilde B} = {\tilde
Q}_1 \cdots {\tilde Q}_n$) have a quantum-deformed moduli space such that
\begin{equation}
\det M - B {\tilde B} = \Lambda^{2N} \ , \label{uni5}
\end{equation}
where $\Lambda^{2N} = \exp(- 8 \pi^2 / g^2)$ is the dynamical scale
of the $SU(3)$ gauge theory. As a consequence, the deformation in
(\ref{uni5}) originates only from the one-instanton contribution.

\section{Heterotic dual models}

\subsection{Heterotic $SO(32)$ model}

Due to the freely-acting nature of the type I orbifold, according to
the adiabatic argument \cite{vw} the S-duality between the type I
and the $SO(32)$ heterotic string is expected to be preserved. In
this section we  explicitly construct the heterotic S-dual of the
$SO(32)$ type I model\footnote{We are grateful to M. Bianchi and E.
Kiritsis for helpful discussions and comments on this point.}. The natural
guess is to use the same freely-acting orbifold generators with a
trivial action on the internal gauge degrees of freedom,
consistently with the fact that in its type I dual the action on the
Chan-Paton factors is trivial as well. There is however one
subtlety, already encountered in similar situations and explained in
other examples in \cite{vw}. Modular invariance forces us to change
the geometric freely-orbifold actions (\ref{gen1})- (\ref{gen3})
into a non-geometric one. Let us consider for simplicity one circle
of radius $R$ and one of the geometric shift in (\ref{gen1}) -
(\ref{gen3})
\begin{equation}
X \ \rightarrow \ X \ + \ \pi R \ . \label{het1}
\end{equation}
Our claim is that its S-dual on the heterotic side is the
non-geometric action\footnote{As shown recently \cite{aci},  such
asymmetric shifts in type I models are consistent only if they act in
an even number of coordinates.}
\begin{equation}
X_L \ \rightarrow \ X_L \ + \ \frac{\pi R}{2} \ + \ \frac{\pi
\alpha'}{2 R} \quad , \quad X_R \ \rightarrow \ X_R \ + \
\frac{\pi R}{2} \ - \ \frac{\pi \alpha'}{2 R} \ . \label{het2}
\end{equation}
In order to prove this claim, we use the fermionic formulation of
the sixteen dimensional heterotic gauge lattice, with $16$ complex
fermions. Guided by the type I dual model, we take a trivial
orbifold action on the $16$ gauge fermions. The adiabatic argument
of \cite{vw} allows identification of the orbifold action only in
the large radius limit, where the shift (\ref{het2}) is
indistinguishable from (\ref{het1}). In the twisted sector of the
theory, the masses of the lattice states $(m,n)$ are shifted according to
\begin{equation}
(m,n) \quad \rightarrow \quad  (m+s_1,n + s'_1) \ , \label{het01}
\end{equation}
where $(s_1,s'_1)=(1/2,0)$ for (\ref{het1}) and $(s_1,s'_1)=(1/2,1/2)$
for (\ref{het2}). The Virasoro generators of the left and right
CFT's are
\begin{eqnarray}
&& L_0 \ = \ N \ + \ 2 \times (-\frac{1}{12} - \frac{1}{24} ) + 2
\times (\frac{1}{24} + \frac{1}{12} ) \ , \nonumber \\
&& {\bar L}_0 \ = \ {\tilde N} \ + \ 10 \times (-\frac{1}{12} ) +
2 \times \frac{1}{24} \ , \label{het02}
\end{eqnarray}
where $N$ (${\tilde N}$) contains the oscillator contributions
whereas the other terms are the zero-point energy in the NS sector
from the spacetime and the gauge coordinates. Level-matching in the
twisted sector is then
\begin{equation}
L_0 \ - \ {\bar L}_0 \ = \ N - {\tilde N} + \frac{3}{4} \ = \ - (m
+ s_1) (n + s'_1) \quad {\rm (mod \ 1)} \ . \label{het03}
\end{equation}
This is possible only for $(s_1,s'_1)=(1/2,1/2)$ which therefore
fixes (\ref{het2}) to be the correct choice.
The S-dual of the type I freely-acting $SO(32)$ is then
defined by the modular invariant torus amplitude
\begin{multline}
T=\int_{\cal F} \frac{d^2 \tau}{\tau_2^3
\eta^2\overline{\eta}^2}\frac{1}{4}
\left[(\tau_{oo}+\tau_{og}+\tau_{oh}+
\tau_{of})\Lambda_1\Lambda_2\Lambda_3 + \right. \\
\left.(\tau_{oo}+\tau_{og}-\tau_{oh}-\tau_{of})(-1)^{m_1+n_1}\Lambda_1\left|\frac{4\eta^2}{\vartheta_2^2}\right|^2+
(\tau_{oo}-\tau_{og}+\tau_{oh}-\tau_{of})(-1)^{m_3+n_3}\Lambda_3\left|\frac{4\eta^2}{\vartheta_2^2}\right|^2+\right.\\
\left.(\tau_{oo}-\tau_{og}-\tau_{oh}+\tau_{of})(-1)^{m_2+n_2}\Lambda_2\left|\frac{4\eta^2}{\vartheta_2^2}\right|^2+
(\tau_{go}+\tau_{gg}+\tau_{gh}+\tau_{gf})\Lambda_1^{m_1+\frac{1}{2},n_1+\frac{1}{2}}\left|\frac{4\eta^2}{\vartheta_4^2}\right|^2
+\right.\\
\left.(\tau_{ho}+\tau_{hg}+\tau_{hh}+\tau_{hf})\Lambda_3^{m_3+\frac{1}{2},n_3+\frac{1}{2}}\left|\frac{4\eta^2}{\vartheta_4^2}\right|^2+
(\tau_{fo}+\tau_{fg}+\tau_{fh}+\tau_{ff})\Lambda_2^{m_2+\frac{1}{2},n_2+\frac{1}{2}}\left|\frac{4\eta^2}{\vartheta_4^2}\right|^2 \right.\\
- \left.(\tau_{go}+\tau_{gg}-\tau_{gh}-\tau_{gf})(-1)^{m_1+n_1}\Lambda_1^{m_1+\frac{1}{2},n_1+\frac{1}{2}}
\left|\frac{4\eta^2}{\vartheta_3^2}\right|^2-\right.\\
\left.
-(\tau_{ho}-\tau_{hg}+\tau_{hh}-\tau_{hf})(-1)^{m_3+n_3}\Lambda_3^{m_3+\frac{1}{2},n_3+\frac{1}{2}}\left|\frac{4\eta^2}{\vartheta_3^2}\right|^2-\right.\\
\left.
- (\tau_{fo}-\tau_{fg}-\tau_{fh}+\tau_{ff})(-1)^{m_2+n_2}\Lambda_2^{m_2+\frac{1}{2},n_2+\frac{1}{2}}\left|\frac{4\eta^2}{\vartheta_3^2}\right|^2
\right] \times (\overline{O}_{32}+\overline{S}_{32})\ . \label{het3}
\end{multline}
Indeed, the massless spectrum matches perfectly with its type I
counterpart. Compared to its type I S-dual cousin, the heterotic
model has the same spectrum for the Kaluza-Klein modes, whereas it
has a different spectrum for the winding modes. This is precisely
what is expected from S-duality \cite{witten2}, which maps KK states into KK
states, whereas it maps perturbative winding states into
non-perturbative states in the S-dual theory.

\subsection{Dual heterotic models with orthogonal gauge groups}

In the fermionic formulation, the dual of the type I $SO(n_o)
\otimes SO(n_g) \otimes SO(n_h) \otimes SO(n_f)$,  $n_0 + n_g + n_f
+ n_h = 32 $ can be constructed by splitting the $16$ complex
fermions of the gauge lattice into $n_0/2 +n_g/2 + n_f/2 + n_h/2 $
groups. We then embed the orbifold action into the gauge lattice as
shown in Table \ref{h001}.
\begin{table}[!ht]
\begin{center}
\begin{tabular}{|c||c|c|c|c|}
\hline {\rm orb. actions} & $SO(n_o)$ & $SO(n_g)$ & $SO(n_f)$ & $SO(n_h)$  \\
\hline \hline
$g$ & $+$ & $+$ & $-$ & $-$ \\
\hline
$f$ & $+$ & $-$ & $+$ & $-$ \\
\hline $h$ & $+$ & $-$ & $-$ & $+$ \\
\hline
\end{tabular}
\end{center}
\caption{Orbifold actions in the gauge degrees of freedom in the
fermionic formulation.}\label{h001}
\end{table}

Level matching in this case can be readily worked out with the
result, in the $g$, $f$ and $h$ twisted sectors respectively
\begin{eqnarray}
&& L_0 \ - \ {\bar L}_0 \ = \ N - {\tilde N} - \frac{5}{4} + \
\frac{n_o + n_g}{16} = \ - (m_1 + s_1) (n_1 + s'_1) \quad {\rm
(mod \ 1)} \ ,
\nonumber \\
&& L_0 \ - \ {\bar L}_0 \ = \ N - {\tilde N} - \frac{5}{4} + \
\frac{n_o + n_f}{16} = \ - (m_2 + s_2) (n_2 + s'_2) \quad {\rm
(mod \ 1)} \ ,
\nonumber \\
&& L_0 \ - \ {\bar L}_0 \ = \ N - {\tilde N} - \frac{5}{4} + \
\frac{n_o + n_h}{16} = \ - (m_3 + s_3) (n_3 + s'_3) \quad {\rm
(mod \ 1)} \ . \label{het04}
\end{eqnarray}
The various possibilities are then as follows

\begin{itemize}
\item $n_o + n_g = 8$ \quad (mod $8$) $\rightarrow$ $s_1=s'_1 = 1/2 \ ,$
\item $n_o + n_g = 4$ \quad (mod $8$) $\rightarrow$ $s_1=1/2$ , $s'_1 = 0 \ ,$
\end{itemize}
and similarly for the other pairs $n_o+n_f$, $n_o+n_h$.
It is interesting to notice the restrictions on the rank of the gauge
group. While the restriction on the even $SO(2n)$ gauge factors was
expected from the beginning, the above conditions are actually
stronger. \\
 Let us take a closer look to the particular case of the gauge group $SO(p)
\otimes SO(q)$ with $p + q= 32$, in order to better understand this
point. The corresponding setting is $n_o=p$, $n_g = q$ and $n_f=n_h=0$.
 Level matching in the $f$ and $h$ twisted sectors reads
\begin{equation}
L_0 \ - \ {\bar L}_0 \ = \ N - {\tilde N} - \frac{5}{4}  + \
\frac{p}{16} = \ - (m + s_1) (n + s_2) \quad {\rm (mod \ 1)} \ ,
\label{het05}
\end{equation}
which leads to the following options:
\begin{itemize}
\item $p = 8$ \quad (mod $8$) $\rightarrow$ $s_1=s_2 = 1/2 \ , $
\item $p = 4$ \quad (mod $8$) $\rightarrow$ $s_1=1/2$ , $s_2 = 0 \ .$
\end{itemize}
Surprisingly, we do not find solutions for $p = 2$ (mod 2). We can only
speculate that, perhaps, a more subtle orbifold actions on the gauge
lattice and/or the introduction of discrete Wilson lines could help in finding
the $p=2$ models, which the dual type I models
suggest that have to exist. \\
For the first case, $p=8,16,24,$ it is convenient, in the fermionic
formulation of the gauge degrees of freedom, to define the following characters
\begin{eqnarray}
&& \chi_o \ = \  O_p O_q \ + \ C_p C_q \quad , \quad \chi_v \ = \
V_p
V_q \ + \ S_p S_q \ , \nonumber \\
&& \chi_s \ = \  O_p C_q \ + \ C_p O_q \quad , \quad \chi_c \ = \
V_p S_q \ + \ S_p V_q  \ .\label{het06}
\end{eqnarray}
The complete partition function of the heterotic model is then
\begin{multline}
T=\int_{\cal F} \frac{d^2 \tau}{\tau_2^3
\eta^2\overline{\eta}^2}\frac{1}{4} \{
\left[(\tau_{oo}+\tau_{og}+\tau_{oh}+
\tau_{of})\Lambda_1\Lambda_2\Lambda_3+\right. \\
\left.+
(\tau_{oo}+\tau_{og}-\tau_{oh}-\tau_{of})(-1)^{m_1+n_1}\Lambda_1
\left|\frac{4\eta^2}{\vartheta_2^2}\right|^2 +
(\tau_{go}+\tau_{gg}+\tau_{gh}+\tau_{gf})\Lambda_1^{m_1+\frac{1}{2},n_1+\frac{1}{2}}\left|\frac{4\eta^2}{\vartheta_4^2}\right|^2
+\right. \\
\left.+(\tau_{go}+\tau_{gg}-\tau_{gh}-\tau_{gf})(-1)^{m_1+n_1}
\Lambda_1^{m_1+\frac{1}{2},n_1+\frac{1}{2}}\left|\frac{4\eta^2}{\vartheta_3^2}\right|^2
\right] (\overline{\chi_o + \chi_v})+ \\
\left[
(\tau_{oo}-\tau_{og}+\tau_{oh}-\tau_{of})(-1)^{m_3+n_3}\Lambda_3
+(\tau_{oo}-\tau_{og}-\tau_{oh}+\tau_{of})(-1)^{m_2+n_2}\Lambda_2
\right] \left|\frac{4\eta^2}{\vartheta_2^2}\right|^2 (\overline{\chi_o - \chi_v}) \\
+ \left[(\tau_{ho}+\tau_{hg}+\tau_{hh}+\tau_{hf})\Lambda_3^{m_3+\frac{1}{2},n_3+\frac{1}{2}}
+ (\tau_{fo}+\tau_{fg}+\tau_{fh}+\tau_{ff})
\Lambda_2^{m_2+\frac{1}{2},n_2+\frac{1}{2}} \right]
\left|\frac{4\eta^2}{\vartheta_4^2}\right|^2 (\overline{\chi_s + \chi_c}) \\
- (-1)^{q / 8}
\left[(\tau_{ho}-\tau_{hg}+\tau_{hh}-\tau_{hf})(-1)^{m_3+n_3}
\Lambda_3^{m_3+\frac{1}{2},n_3+\frac{1}{2}}+
 \right.\\
\left.+ (\tau_{fo}-\tau_{fg}-\tau_{fh}+\tau_{ff})(-1)^{m_2+n_2}
\Lambda_2^{m_2+\frac{1}{2},n_2+\frac{1}{2}} \right]
\left|\frac{4\eta^2}{\vartheta_3^2}\right|^2
 (\overline{\chi_s - \chi_c}) \}  \ . \label{het07}
\end{multline}
As for the $SO(32)$ model, the whole KK spectrum precisely match
the corresponding one on the type I S-dual side, whereas the massive winding states and the massive
twisted spectra are, as expected, quite different. \\
On the other hand, for the second case $p=4,12,20$, the correct
characters are
\begin{eqnarray}
&& \chi_o \ = \  O_p O_q \ + \ C_p C_q \quad , \quad \chi_v \ = \
V_p
V_q \ + \ S_p S_q \nonumber \\
&& \chi_s \ = \  V_p C_q \ + \ S_p O_q \quad , \quad \chi_c \ = \
O_p S_q \ + \ C_p V_q \label{het08}
\end{eqnarray}
The complete partition function is now
\begin{multline}
T=\int_{\cal F} \frac{d^2 \tau}{\tau_2^3
\eta^2\overline{\eta}^2}\frac{1}{4} \{
\left[(\tau_{oo}+\tau_{og}+\tau_{oh}+
\tau_{of})\Lambda_1\Lambda_2\Lambda_3+\right. \\
\left.+
(\tau_{oo}+\tau_{og}-\tau_{oh}-\tau_{of})(-1)^{m_1+n_1}\Lambda_1
\left|\frac{4\eta^2}{\vartheta_2^2}\right|^2 +
(\tau_{go}+\tau_{gg}+\tau_{gh}+\tau_{gf})\Lambda_1^{m_1+\frac{1}{2},n_1+\frac{1}{2}}\left|\frac{4\eta^2}{\vartheta_4^2}\right|^2
+\right. \\
\left.+(\tau_{go}+\tau_{gg}-\tau_{gh}-\tau_{gf})(-1)^{m_1+n_1}
\Lambda_1^{m_1+\frac{1}{2},n_1+\frac{1}{2}}\left|\frac{4\eta^2}{\vartheta_3^2}\right|^2
\right] (\overline{\chi_o + \chi_v})+ \\
\left[
(\tau_{oo}-\tau_{og}+\tau_{oh}-\tau_{of})(-1)^{m_3}\Lambda_3
+(\tau_{oo}-\tau_{og}-\tau_{oh}+\tau_{of})(-1)^{m_2}\Lambda_2
\right] \left|\frac{4\eta^2}{\vartheta_2^2}\right|^2 (\overline{\chi_o - \chi_v}) \\
+ \left[(\tau_{ho}+\tau_{hg}+\tau_{hh}+\tau_{hf})\Lambda_3^{m_3,n_3+\frac{1}{2}}
+ (\tau_{fo}+\tau_{fg}+\tau_{fh}+\tau_{ff})
\Lambda_2^{m_2,n_2+\frac{1}{2}} \right]
\left|\frac{4\eta^2}{\vartheta_4^2}\right|^2 (\overline{\chi_s + \chi_c}) \\
- (-1)^{(p+4) / 8}
\left[(\tau_{ho}-\tau_{hg}+\tau_{hh}-\tau_{hf})(-1)^{m_3}
\Lambda_3^{m_3,n_3+\frac{1}{2}} \right.\\
\left.+ (\tau_{fo}-\tau_{fg}-\tau_{fh}+\tau_{ff})(-1)^{m_2}
\Lambda_2^{m_2,n_2+\frac{1}{2}} \right]
\left|\frac{4\eta^2}{\vartheta_3^2}\right|^2
 (\overline{\chi_s - \chi_c}) \}  \ . \label{het09}
\end{multline}
It should be noticed that while the  KK spectra are actually the same for the two cases $p=4$
and $p=8$ (mod 8), they are very different in the massive winding
sector, in perfect agreement with the modular invariance constraints
({\ref{het04}}).

We expect that appropriate orbifold action in the sixteen dimensional gauge
lattice will also produce the S-dual of the type I racetrack and
of the unitary gauge group cases, discussed in the previous
sections. The required action, however, cannot correspond
to a standard Wilson line in the adjoint of the gauge group, but rather to
a non-diagonal action in the Cartan basis, like the ones considered in \cite{nonabelianw}.

\section{Threshold corrections to the gauge couplings}

In this section we perform the one-loop calculation of the threshold
corrections to the gauge couplings of some of the models described in
the previous sections. The effective field theory quantities can be then
easily extracted from the one-loop computation.

The threshold correction $\Lambda_2$ is
generically written as
\begin{equation}
\label{thresgeneral}
\left.\frac{4\pi^2}{g_{a}^2}\right|_{\textrm{1-loop}}= \ \left.\frac{4\pi^2}{g_a^2}\right|_{\textrm{tree}}+ \ \Lambda_{2,a}\ ,
\end{equation}
with
\begin{equation}
\Lambda_{2,a} = \int_{\mathcal{F}} \frac{d^2\tau}{4 \tau_2} \ {\cal B}_a (\tau) \quad
\end{equation}
for the heterotic string, and
\begin{equation}
\Lambda_{2,a} = \int_{0}^{\infty} \frac{dt}{4 t} \ {\cal B}_a (t) \quad
\end{equation}
for the type I string.
In these expressions, ${\cal B}_a$ flows in the infrared to
\begin{equation}
b_a \ = \ - 3 T_a (G) + \sum_r T_a (r) \quad  , \label{thre1}
\end{equation}
the one-loop beta function for the gauge group factor $G_a$, with $r$ running over the
gauge group representations with Dynkin index $T_a (r)$. From
the one-loop expression of the gauge coupling it is possible to
extract \cite{abd} the holomorphic gauge couplings $f_a (M_i)$,
where $M_i$ denote here collectively the moduli chiral
(super)fields, using the relation \cite{kl}
\begin{equation}
\frac{4 \pi^2}{g_a^2 (\mu^2)} \ = \ Re f_a \ + \ \frac{b_a}{4} \
\textrm{log} \ \frac{M_P^2}{\mu^2} + \frac{c_a}{4} K + \frac{T_a (G)}{2}
\ln g_a^{-2} (\mu^2) - \sum_r \frac{T_a (r)}{2} \ln \textrm{det } Z_r
(\mu)^2 \ , \label{thre2}
\end{equation}
where $K$ is the Kahler potential, $Z_r$ is the wave-function
normalization matrix for the matter fields and $c_a = \sum_r T_a (r) - T_a (G)$. With this
definition, the holomorphic non-perturbative scale $\Lambda_a$ of an
asymptotically-free gauge theory ($b_a < 0 $) is given by
\begin{equation}
\Lambda_a \ = \ M_P \ e^{- \frac{2 f_a}{|b_a|}} \ . \label{thre3}
\end{equation}


\subsection{Type I $SO(n_o) \otimes SO(n_g) \otimes SO(n_f) \otimes SO(n_h)$ model}

For the computation of threshold corrections to the gauge couplings in
the freely-acting type I model with orthogonal gauge groups, we make use of
the background field method~\cite{bachasporrati,bachasfabre,abd}. Therefore, we
introduce a magnetic field along two of the spatial non-compact
directions, say $F_{23}=BQ$. In the weak field limit, the one-loop
vacuum energy can be expanded in powers of $B$, providing
\begin{equation}
\Lambda(B)=\Lambda_0+\frac{1}{2}\left(\frac{B}{2\pi}\right)^2\Lambda_2+\ldots\ .
\end{equation}
For supersymmetric vacua $\Lambda_0=0$, and the quadratic term accounts exactly for the threshold corrections
in eq.(\ref{thresgeneral}).

In the presence of $F_{23}$, the oscillator
modes along the non-compact complex plane $x^2+ix^3$ get
shifted by an amount $\epsilon$ such that
\begin{equation}
\pi\epsilon = \textrm{arctan}(\pi q_LB)+\textrm{arctan}(\pi q_R B)\simeq \pi (q_L+q_R)B + O(B^3)\ ,
\label{epsmode}
\end{equation}
where $q_L$ and $q_R$ are the eigenvalues of the gauge group
generator $Q$, acting on the Chan-Paton states localized at the two
endpoints of the open strings. In the vacuum energy, the contribution
of the non-compact bosons and fermions gets replaced by
\begin{equation}
\frac{\vartheta_{\alpha}(0|\tau)}{\eta^3(\tau)}\ \to \ 2 \pi \epsilon
\tau
\frac{\vartheta_{\alpha}(\tau\epsilon|\tau)}{\vartheta_1(\tau\epsilon|\tau)}
\quad \quad \textrm{for }\alpha=2,3,4 \
\end{equation}
in the annulus and M\"obius amplitudes.  In addition, the momentum operator along the non-compact
dimensions becomes,
\begin{equation}
p^{\mu}p_{\mu} \ \to \ -(p_0)^2+(p_1)^2+(2n+1)\epsilon + 2\epsilon\Sigma_{23}\ ,
\end{equation}
where $\Sigma_{23}$ is the spin operator in the $(23)$ direction, while
$n$ is an integer that labels the Landau levels.  The
supertrace operator becomes now
\begin{equation}
\textrm{STr} \ \to \ \left(\sum_{bos}-\sum_{ferm}\right)\frac{(q_L+q_R)B}{2\pi}\int\frac{d^2p}{(2\pi)^2}\ ,
\end{equation}
where $(q_L+q_R)B/2\pi$ is the density of the Landau levels and the integral
is performed only over the momenta in
the non-compact directions $x^0$ and $x^1$.

The details of the computation can be found in Appendix \ref{appiso}.
Collecting the results obtained there, and assuming $Q$ to be in a $U(1)$ inside
$SO(n_o)$, $SO(n_g)$, $SO(n_f)$ or $SO(n_h)$,
the moduli dependent threshold corrections for the respective gauge couplings can be written
as follows,
\begin{eqnarray}
\Lambda_{2,o}&=-\frac{1}{4} \textrm{Tr}(Q^2)\left[(2-g_N)\left(\pi
    \textrm{Re }U_1+\textrm{log}[(\textrm{Re }U_1)(\textrm{Re }T_1)
    \mu^2 \left|\frac{\vartheta_4}{\eta^3}(2iU_1)\right|^{-2} ] \right)+\right. \nonumber \\
&\left.+(2-f_N)\left(\pi \textrm{Re }U_2+\textrm{log}[(\textrm{Re
    }U_2)(\textrm{Re }T_2) \mu^2  \left|\frac{\vartheta_4}{\eta^3}(2iU_2)\right|^{-2} ] \right)+\nonumber \right. \\
&\left.+(2-h_N)\left(\pi \textrm{Re }U_3+\textrm{log}[(\textrm{Re
    }U_3)(\textrm{Re }T_3) \mu^2 \left|\frac{\vartheta_4}{\eta^3}(2iU_3)\right|^{-2} ] \right)\right]\ ,
\end{eqnarray}
\begin{eqnarray}
\Lambda_{2,g}&=-\frac{1}{4} \textrm{Tr}(Q^2)\left[(2-g_N)\left(\pi
    \textrm{Re }U_1+\textrm{log}[(\textrm{Re }U_1)(\textrm{Re }T_1)
    \mu^2 \left|\frac{\vartheta_4}{\eta^3}(2iU_1)\right|^{-2} ] \right)+\right. \nonumber \\
&\left.+(2+f_N)\left(\pi \textrm{Re }U_2+\textrm{log}[(\textrm{Re
    }U_2)(\textrm{Re }T_2) \mu^2 \left|\frac{\vartheta_4}{\eta^3}(2iU_2)\right|^{-2} ] \right)+\nonumber \right. \\
&\left.+(2+h_N)\left(\pi \textrm{Re }U_3+\textrm{log}[(\textrm{Re
    }U_3)(\textrm{Re }T_3) \mu^2 \left|\frac{\vartheta_4}{\eta^3}(2iU_3)\right|^{-2} ] \right)\right]\ ,
\end{eqnarray}
\begin{eqnarray}
\Lambda_{2,f}&=-\frac{1}{4} \textrm{Tr}(Q^2)\left[(2+g_N)\left(\pi
    \textrm{Re }U_1+\textrm{log}[(\textrm{Re }U_1)(\textrm{Re }T_1)
    \mu^2 \left|\frac{\vartheta_4}{\eta^3}(2iU_1)\right|^{-2} ] \right)+\right. \nonumber \\
&\left.+(2-f_N)\left(\pi \textrm{Re }U_2+\textrm{log}[(\textrm{Re
    }U_2)(\textrm{Re }T_2) \mu^2 \left|\frac{\vartheta_4}{\eta^3}(2iU_2)\right|^{-2} ] \right)+\nonumber \right. \\
&\left.+(2+h_N)\left(\pi \textrm{Re }U_3+\textrm{log}[(\textrm{Re
    }U_3)(\textrm{Re }T_3) \mu^2 \left|\frac{\vartheta_4}{\eta^3}(2iU_3)\right|^{-2} ] \right)\right]\ ,
\end{eqnarray}
\begin{eqnarray}
\Lambda_{2,h}&=-\frac{1}{4} \textrm{Tr}(Q^2)\left[(2+g_N)\left(\pi
    \textrm{Re }U_1+\textrm{log}[(\textrm{Re }U_1)(\textrm{Re }T_1)
    \mu^2 \left|\frac{\vartheta_4}{\eta^3}(2iU_1)\right|^{-2} ] \right)+\right. \nonumber \\
&\left.+(2+f_N)\left(\pi \textrm{Re }U_2+\textrm{log}[(\textrm{Re
    }U_2)(\textrm{Re }T_2) \mu^2
    \left|\frac{\vartheta_4}{\eta^3}(2iU_2)\right|^{-2} ] \right)+\nonumber \right. \\
&\left.+(2-h_N)\left(\pi \textrm{Re }U_3+\textrm{log}[(\textrm{Re
    }U_3)(\textrm{Re }T_3) \mu^2
    \left|\frac{\vartheta_4}{\eta^3}(2iU_3)\right|^{-2} ]
  \right) \right] \ .
\end{eqnarray}
The $\beta$-function coefficients can also be extracted in the form
\begin{align}
b_o& = - [3(n_o-2)-(n_f+n_g+n_h)]\ , \nonumber \\
b_g&=  - [3(n_g-2)-(n_f+n_o+n_h)]\ , \nonumber \\
b_f&=  - [3(n_f-2)-(n_o+n_g+n_h)]\ , \nonumber \\
b_h&=  - [3(n_h-2)-(n_f+n_g+n_o)]\ ,
\end{align}
and, using the definition (\ref{thre2}), the holomorphic one-loop gauge kinetic functions
are then
\begin{align}
f_o& =  S + \frac{1}{2} \left[(2-g_N) \textrm{log} \frac{\vartheta_4}{e^{\pi U_1/2} \eta^3} (2 i
U_1) +(2-f_N) \textrm{log} \frac{\vartheta_4}{e^{\pi U_2/2} \eta^3} (2 i U_2) +\right. \nonumber \\
&\left. \qquad +(2-h_N) \textrm{log} \frac{\vartheta_4}{e^{\pi U_3/2} \eta^3} (2 i U_3) \right]   \ , \nonumber
\end{align}
\begin{align}
f_g& =  S + \frac{1}{2} \left[(2-g_N) \textrm{log} \frac{\vartheta_4}{e^{\pi U_1/2} \eta^3} (2 i
U_1) +(2+f_N) \textrm{log} \frac{\vartheta_4}{e^{\pi U_2/2} \eta^3} (2 i U_2) +\right. \nonumber \\
&\left.\qquad +(2+h_N) \textrm{log} \frac{\vartheta_4}{e^{\pi U_3/2} \eta^3} (2 i U_3) \right] \ , \nonumber
\end{align}
\begin{align}
f_f& =  S + \frac{1}{2} \left[(2+g_N) \textrm{log} \frac{\vartheta_4}{e^{\pi U_1/2} \eta^3} (2 i
U_1) +(2-f_N) \textrm{log} \frac{\vartheta_4}{e^{\pi U_2/2} \eta^3} (2 i U_2) +\right. \nonumber \\
&\left.\qquad +(2+h_N) \textrm{log} \frac{\vartheta_4}{e^{\pi U_3/2} \eta^3} (2 i U_3) \right] \ , \nonumber
\end{align}
\begin{align}
f_h& =  S + \frac{1}{2} \left[(2+g_N) \textrm{log} \frac{\vartheta_4}{e^{\pi U_1/2} \eta^3} (2 i
U_1) +(2+f_N) \textrm{log} \frac{\vartheta_4}{e^{\pi U_2/2} \eta^3} (2 i U_2) +\right. \nonumber \\
&\left.\qquad +(2-h_N) \textrm{log} \frac{\vartheta_4}{e^{\pi U_3/2}  \eta^3} (2 i U_3) \right]
\ . \label{thref}
\end{align}
It is very important to stress the linear dependence of the above
threshold corrections on the ($\pi \textrm{Re }U_i$) factors.  Indeed, the presence of such
terms in a loop contribution may seem surprising.  However, expanding the factor $\vartheta_4
\eta^{-3}$, it can be realized that this term exactly cancels the
contributions coming from the factor $q^{1/24}$ contained in the
$\eta$-function. Thus, the total dependence on the moduli of the
threshold corrections turns out to be exclusively of
logarithmic form. This phenomenon can be physically understood making the
observation that, beyond the Kaluza-Klein scale, ${\cal N}=4$ supersymmetry
is effectively recovered. Therefore, in the large volume limit
only logarithmic corrections in the moduli should be present. The
price one has to pay is that modular invariance in the target space is lost,
as evident from the above expressions.  The breaking of modular
invariance in the target space by the shift $\mathbb{Z}_2\times \mathbb{Z}_2$
orbifold is very different from what happens in the ordinary
$\mathbb{Z}_2\times \mathbb{Z}_2$ case where, beyond the Kaluza-Klein
scale, the effective supersymmetry for each sector is still ${\cal N}=2$.
The threshold corrections in that case turn out to be proportional to
$(\textrm{Re }U)\textrm{log}|\eta(iU)|^4$. Therefore, they preserve
modular invariance, but have a non-logarithmic dependence on
the moduli, due to the term $q^{1/24}$ inside the $\eta$-function.

\subsection{Type I racetrack model}

The details of the calculation can be found again in Appendix \ref{apprac}.
Using the background field method, the moduli dependent
part of the gauge coupling threshold corrections is given by
\begin{multline}
\Lambda_{2,p}=- \frac{1}{4} \textrm{Tr}(Q^2)\left[(2-p)\sum_{j=1}^3\left(\pi
    \textrm{Re }U_j+\textrm{log}[(\textrm{Re }U_j)(\textrm{Re }T_j)
    \mu^2 \left|\frac{\vartheta_4}{\eta^3}(2iU_j)\right|^{-2} ] \right)-\right.\\
\left.+q\left(\textrm{log}\left|\frac{\vartheta_4}{\eta^3}(2iU_1)\right|^2-\textrm{log}\left|\frac{\vartheta_4}{\eta^3}(4iU_1)\right|^2+\pi \textrm{Re }U_1\right)\right]\ \ ,
\end{multline}
together with a similar expression for the $SO(q)$ factor, with the obvious replacements.  The corresponding
$\beta$-function coefficients of the $SO(p)$
and $SO(q)$ gauge group factors are
\begin{equation}
b_p \ = \  - \ 3 \ (p-2) \quad , \quad b_q \ = \  - \ 3 \ (q-2) \ , \label{threrace1}
\end{equation}
and the one-loop holomorphic gauge functions read
\begin{align}
& f_p \ = \ S \ + \ \frac{2-p}{2} \sum_{i=1}^3 \textrm{log} \frac{\vartheta_4}{e^{\pi U_i/2}  \eta^3} (2 i U_i) - \frac{q}{2} \left[ \textrm{log}  \frac{\vartheta_4}{e^{\pi U_1/2}  \eta^3} (2 i U_1) - \textrm{log}  \frac{\vartheta_4}{e^{\pi U_1} \eta^3} (4 i U_1) \right] \
, \nonumber \\
&  f_q \ = \ S \ + \ \frac{2-q}{2} \sum_{i=1}^3 \textrm{log} \frac{\vartheta_4}{e^{\pi U_i/2}  \eta^3} (2 i U_i) - \frac{p}{2} \left[ \textrm{log}  \frac{\vartheta_4}{e^{\pi U_1/2} \eta^3} (2 i U_1) - \textrm{log}  \frac{\vartheta_4}{e^{\pi U_1}
 \eta^3} (4 i U_1) \right] \
. \nonumber
\end{align}
The non-perturbative superpotential can be written, in analogy with
(\ref{race2}),
\begin{equation}
W_{np} \ = \ A_p (U_i) \ e^{- a_p S} \ + \ A_q (U_i) \ e^{- a_q S} \
, \label{threrace3}
\end{equation}
where
\begin{align}
a_p & = \frac{2}{p-2} & , \quad A_p & =  \left[ \prod_{i=1}^3
e^{-\pi U_i/2} \frac{\vartheta_4}{\eta^3} (2 i U_i) \right]
\left[e^{\pi U_1/2} \frac{\vartheta_4}{\eta^3} (2 i U_1)
\frac{\eta^3}{\vartheta_4} (4 i U_1) \right]^{\frac{q}{p-2}} \ , \
\nonumber
 \\
a_q & =  \frac{2}{q-2} & , \quad  A_q & =  \left[ \prod_{i=1}^3
e^{-\pi U_i/2} \frac{\vartheta_4}{\eta^3} (2 i U_i) \right]
\left[e^{\pi U_1/2} \frac{\vartheta_4}{\eta^3} (2 i U_1)
\frac{\eta^3}{\vartheta_4} (4 i U_1) \right]^{\frac{p}{q-2}} \ .
\label{threrace4}
\end{align}

\subsection{Heterotic $SO(32)$ model}
\label{hetthresec}

For the heterotic string, several procedures are available in literature
to extract the threshold corrections \cite{kaplu,dkl,kk2}. The general expression for the threshold corrections
to the gauge couplings, valid in the
$\overline{\textrm{DR}}$ renormalization scheme, is given by
\begin{equation}
\Lambda_{2,a}
=\int_{\mathcal{F}}\frac{d^2\tau}{\tau_2}\frac{i}{4\pi}\frac{1}{|\eta|^2}\sum_{\alpha
  ,\beta=0,1/2}
\partial_\tau\left(\frac{\vartheta \left[{\alpha \atop \beta }\right]}{\eta}\right)
\left(Q^2_a-\frac{1}{4\pi\tau_2}\right)C\left[{\alpha \atop \beta }\right]\ , \label{threskap}
\end{equation}
where $Q_a$ is the charge operator of the gauge group $G_a$, and
$C\left[{\alpha \atop \beta }\right]$ is the internal six-dimensional
partition function, which, for the particular case of the $SO(32)$
model, can be read from (\ref{het3}). As noticed in \cite{dkl},
only the ${\cal N}=2$ sectors of the theory contribute to the
moduli dependent part of this expression.

Again, the details of the computation are relegated to Appendix \ref{apphet}.  The expression for
the gauge threshold corrections of the heterotic $SO(32)$ model is
\begin{multline}
\Lambda_{2}=-\frac{1}{96}\int_{\mathcal{F}}\frac{d^2\tau}{\tau_2}\sum_{i=1}^3\left[
(-1)^{m_i+n_i}\hat
Z_i\overline{\vartheta}_3^2\overline{\vartheta}_4^2-
\hat Z_i^{m_i+\frac{1}{2},n_i+\frac{1}{2}}\overline{\vartheta}_2^2\overline{\vartheta}_3^2-\right. \\
\left.-(-1)^{m_i+n_i}\hat
Z_i^{m_i+\frac{1}{2},n_i+\frac{1}{2}}\overline{\vartheta}_2^2\overline{\vartheta}_4^2\right]\frac{\overline{E}_4(\overline{\hat
E}_2\overline{E}_4-\overline{E}_6)}{\bar{\eta}^{24}}\ ,
\label{threhet}
\end{multline}
where $E_{2n}$ are the Eisenstein series (given explicitly in the Appendix \ref{formulae}), and
the three toroidal lattice sums, $\hat Z_i\equiv |\eta|^4\Lambda_i$, read
\begin{equation}
\hat{\mathbf{Z}}_i\left[{h \atop g}\right] = \frac{\textrm{Re }T_i}{\tau_2}\sum_{n_1,\ell^1,n_2,\ell^2}(-1)^{h n_1 + g \ell_1}\textrm{exp}\left[2\pi T_i
\textrm{det}(A)-\frac{\pi (\textrm{Re }T_i)}{\tau_2(\textrm{Re }U_i)}\left|\begin{pmatrix}1&iU_i\end{pmatrix}A\begin{pmatrix}\tau\\ 1\end{pmatrix}\right|^2\right]\ , \label{lat11}
\end{equation}
with
\begin{equation}
A=\begin{pmatrix}n_1+\frac{g}{2}& \ell_1+\frac{h}{2}\label{lat13} \\
n_2& \ell_2\end{pmatrix}
\end{equation}
and
\begin{equation}
(-1)^{m_i+n_i}\hat Z_i=\hat{\mathbf{Z}}_i\left[{1 \atop 0}\right]\ , \quad \hat Z_i^{m_i+\frac{1}{2},n_i+\frac{1}{2}} = \hat{\mathbf{Z}}_i\left[{0 \atop 1}\right]\ , \quad (-1)^{m_i+n_i}\hat Z_i^{m_i+\frac{1}{2},n_i+\frac{1}{2}}=\hat{\mathbf{Z}}_i\left[{1 \atop 1}\right]\ .\nonumber
\end{equation}
Notice that $\left[{h \atop g}\right]_i$ labels the three
$\mathcal{N}=2$ sectors associated to the $i$-th 2-torus,
$i=1,2,3$. Although the full expression
(\ref{threhet}) is worldsheet modular invariant, each of these
$\mathcal{N}=2$ sectors is not worldsheet modular invariant by
itself, contrary to what happens in orbifolds with a
trivial action on the winding modes.

In the large volume limit, $\textrm{Re }T_i \gg 1$, the winding modes decouple and
only Kaluza-Klein modes with small $q$ contribute to the integral.
In that case, the threshold correction receives contributions only
from $A$ matrices with zero determinant in the sector $(h, g)=(1,0)$, in such a way that
(\ref{threhet}) becomes\footnote{We have neglected an extra term coming from
the non-holomorphic regularization of $\hat E_2$, which in the dual type I
side would presumably correspond to contact contributions in two-loop open string diagrams.}
\begin{equation}
\Lambda_2|_{\textrm{Re }T_i\gg 1} \simeq \frac{b}{3}\left[-\pi \textrm{Re }U_i-\textrm{log}[(\textrm{Re }U_i)(\textrm{Re }T_i)\left|\frac{\eta^3}{\vartheta_4}(2iU_i)\right|^2\mu^2]\right]\ ,
\end{equation}
matching exactly the threshold corrections for the dual type I SO(32) model.

For arbitrary $T_i$, however, the winding modes do not decouple from the low
energy physics and  corrections due to worldsheet instantons appear:
\begin{equation}
\Lambda_2 \simeq \Lambda_2|_{\textrm{Re }T_i\gg 1}(U_i) + \Lambda_{\textrm{inst.}}(U_i,T_i) \ . \label{npcor}
\end{equation}
They correspond to $E1$ instanton contributions in the dual type
I SO(32) model, and therefore are absent in (\ref{thref}).

For example, consider the $q\to 0$ contributions to $\Lambda_{\textrm{inst.}}$
of winding modes in the sector $(h,g)=(1,0)$. These result in
\begin{equation}
\left.\Lambda_{\textrm{inst}, \left[{1 \atop
0}\right]}\right|_{q\to 0}\simeq -\frac{2b}{3}\sum_{n=1}^{\infty}
(-1)^n \textrm{log}\prod_{i=1}^3(1-e^{-2\pi nT_i})\ + \
\textrm{c.c.}
\end{equation}
Since the axionic part of $T_i$ in type I corresponds to
components of the RR 2-form, $C_2$, it is natural to expect that these
contributions come from $E1$ instantons wrapping $n$ times the
(1,1)-cycle associated to $T_i$. Notice that the dependence on
$T_i$ perfectly agrees with general arguments in \cite{germans}
for the mirror type IIA picture.

The corresponding holomorphic gauge kinetic function reads
\begin{equation}
f = S - 15 \sum_{i=1}^3 \left[  \textrm{log}
  \frac{\vartheta_4}{e^{\pi U_i/2} \eta^3}(2iU_i) - 2 \ \sum_{n=1}^{\infty}
  (-1)^n \textrm{log} (1 - e^{- 2 \pi n T_i}) \right]+\ldots  \ \ , \label{threhet4}
\end{equation}
where the dots denote further contributions from $\Lambda_{\textrm{inst.}}$.
Hence, the non-perturbative superpotential generated by
gaugino condensation receives an extra dependence in the K\"ahler
moduli,
\begin{equation}
W_{np} \ = \ A (U_i,T_i) \ e^{- a S} \ , \label{threhet5}
\end{equation}
with
\begin{equation}
a = \frac{1}{15} \quad , \quad A =   \prod_{i=1}^3 \left[ e^{-\pi
U_i/2} \frac{\vartheta_4}{\eta^3} (2 i U_i) \prod_{n=1}^{\infty} \left(
\frac{1 - e^{- 4 \pi (n + 1/2) T_i}}{1 - e^{- 4 \pi n  T_i}} \right)^2 \right]\times \ldots \ \ .
\label{threhet6}
\end{equation}
Unfortunately, a complete analytic evaluation of the non-perturbative
corrections in (\ref{threhet}) is subtle, as
worldsheet modular invariance mix orbits within different
$\mathcal{N}=2$ sectors and the
unfolding techniques of \cite{dkl,bfkov} cannot be applied
straightforwardly to this case.

\section{Euclidean brane instantons in the type I freely-acting
$SO(32)$ model}

The model has two types of BPS brane instantons, denoted as $E5$ and $E1$. The
$E5$ branes are interpreted as gauge instantons within the four dimensional gauge
theory on the compactified $D9$ branes and map, in the heterotic
dual, to non-perturbative euclidean NS5 corrections. The $E1_i$
type I instantons wrapping the internal torus $T^i$, instead, are stringy
instantons from the gauge theory perspective and are responsible,
in the heterotic dual, for the perturbative world-sheet instantons
effects, that we have computed in section 5.\footnote{Notice that generically there will be also massless modes stretching
between both kind of instantons, $E5$ and $E1_i$. From the gauge theory perspective, these
modes are presumably responsible of the $E1$ instanton corrections to the Veneziano-Yankielowicz
superpotential, discussed at the end of section \ref{hetthresec}.}

The configurations of the various $Op$ planes and $(D/E)p$ branes in the models are
pictorially provided in table \ref{e01}.

\begin{table}[!ht]
\begin{center}
\begin{tabular}{|c||c|c|c|c|c|c|c|c|c|c|}
\hline
{\rm coord.} & 0 & 1 & 2 & 3 & 4 & 5 & 6 & 7 &
8 & 9 \\
\hline \hline
$D9/O9$ & $-$ & $-$ & $-$ & $-$ & $-$ & $-$ & $-$ & $-$ & $-$ & $-$ \\
\hline
$O5_1$ & $-$ & $-$ & $-$ & $-$ & $-$ & $-$ & $\bullet$ & $\bullet$ & $\bullet$ & $\bullet$ \\
\hline
$O5_2$ & $-$ & $-$ & $-$ & $-$ & $\bullet$ & $\bullet$ & $-$ & $-$ & $\bullet$ & $\bullet$  \\
\hline
$O5_3$ & $-$ & $-$ & $-$ & $-$ & $\bullet$ & $\bullet$ & $\bullet$ & $\bullet$ & $-$ & $-$  \\
\hline
$E1_1$ & $\bullet$ & $\bullet$ & $\bullet$ & $\bullet$ & $-$ & $-$ & $\bullet$ & $\bullet$ & $\bullet$ & $\bullet$  \\
\hline
$E1_2$ & $\bullet$ & $\bullet$ & $\bullet$ & $\bullet$ & $\bullet$ & $\bullet$ & $-$ & $-$ & $\bullet$ & $\bullet$  \\
\hline
$E1_3$ & $\bullet$ & $\bullet$ & $\bullet$ & $\bullet$ & $\bullet$ & $\bullet$ & $\bullet$ & $\bullet$ & $-$ & $-$  \\
\hline
$E5$   & $\bullet$ & $\bullet$ & $\bullet$ & $\bullet$ & $-$ & $-$ & $-$ & $-$ & $-$ & $-$ \\
\hline
\end{tabular}
\end{center}
\caption{$Op$-planes and $D9/Ep$ branes present in the type I models. A  - denotes a
coordinate parallel to the $Op$-plane/$Dp$-brane, while a  $\bullet$ represent an orthogonal
coordinate.}\label{e01}
\end{table}
\subsection{E5 instantons}

A convenient way to describe the $E5$ instantons is to write the
partition functions coming from the cylinder amplitudes (for $E5$-$E5$ and $E5$-$D9$
strings) and the M\"obius amplitudes (for $E5$-$O9$ and $E5$-$O5_{i}$).  In order to extract the
spectrum, it is useful to express
the result using the subgroup of $SO(10)$ involved in a covariant
description, namely $SO(4) \times SO(2)^3$ in our present case.
Considering $p$ coincident $E5$ instantons, one gets
\begin{align}
A_{E5-E5} &=  \frac{p^2}{16} \int_0^{\infty} \frac{dt}{t} \frac{1}{\eta^2}
\sum_{\alpha \beta} c_{\alpha \beta}
\frac{\vartheta [{\alpha \atop \beta }]}{\eta}
\{  P_1 P_2 P_3 \frac{\vartheta [{\alpha \atop \beta }]^3}{\eta^9}
+ (-1)^{m_1} P_1 \frac{\vartheta [{\alpha \atop \beta }]\vartheta [{\alpha
    \atop \beta+1/2 }] \vartheta [{\alpha \atop \beta-1/2 }]}{\eta^5} \frac{4 \eta^2}{\vartheta_2^2} \nonumber \\
& +
 (-1)^{m_2} P_2 \frac{\vartheta [{\alpha \atop \beta-1/2 }]\vartheta [{\alpha
    \atop \beta }] \vartheta [{\alpha \atop \beta+1/2 }]}{\eta^5} \frac{4 \eta^2}{\vartheta_2^2}  +
  (-1)^{m_3} P_3 \frac{\vartheta [{\alpha \atop \beta +1/2}]\vartheta [{\alpha
    \atop \beta-1/2 }] \vartheta [{\alpha \atop \beta }]}{\eta^5} \frac{4 \eta^2}{\vartheta_2^2}  \} \label{e2}
\end{align}
\begin{align}
M_{E5-O9} &=  - \frac{p}{16} \int_0^{\infty} \frac{dt}{t} \frac{4 \eta^2}{\vartheta_2^2} \eta^2 \sum_{\alpha \beta}  c_{\alpha \beta}
\frac{\vartheta [{\alpha \atop \beta+1/2 }] \vartheta [{\alpha \atop
      \beta-1/2 }]}{\eta^2} \frac{\eta}{\vartheta [{\alpha \atop
      \beta }]}  \{  P_1 P_2 P_3 \frac{\vartheta [{\alpha \atop \beta }]^3}{\eta^9}
+  \nonumber \\
& \left[ (-1)^{m_1} P_1 \frac{\vartheta [{\alpha \atop \beta }]\vartheta [{\alpha
    \atop \beta+1/2 }] \vartheta [{\alpha \atop \beta-1/2 }]}{\eta^3} \!+\!
 (-1)^{m_2} P_2 \frac{\vartheta [{\alpha \atop \beta-1/2 }]\vartheta [{\alpha
    \atop \beta }] \vartheta [{\alpha \atop \beta+1/2 }]}{\eta^3} \!+\right. \nonumber \\
& \left. + (-1)^{m_3} P_3 \frac{\vartheta [{\alpha \atop \beta +1/2}]\vartheta [{\alpha
    \atop \beta-1/2 }] \vartheta [{\alpha \atop \beta }]}{\eta^3} \right] \times \frac{1}{\eta^2} \frac{4\eta^2}{\vartheta_2^2}  \} \ , \label{e3}
\end{align}
where $c_{\alpha \beta}$ are the usual GSO projection coefficients.
In terms of covariant  $SO(4) \times SO(2)^3$ characters, the
massless instanton zero-modes content results
\begin{equation}
A_{E5-E5}^{(0)} + M_{E5-O9}^{(0)} = \frac{p (p+1)}{2} \ (V_4O_2O_2O_2 - C_4
  C_2C_2C_2) \ - \ \frac{p (p-1)}{2} \ S_4 S_2 S_2 S_2 \ . \label{e4}
\end{equation}
From a four-dimensional perspective, $V_4O_2O_2O$ describe vector zero-modes,
$a_{\mu}$, while $C_4 C_2C_2C_2$ is a spinor $M^{\alpha,- - -}$, where
$\alpha$ denotes an $SO(4)$ spinor index of positive chirality, whereas
$(---)$ denote the $SO(2)^3$ internal chiralities. Analogously, $S_4 S_2
S_2 S_2$ are fermionic zero modes $\lambda^{{\dot \alpha},- -
  -}$. Notice that in the one-instanton $p=1$ sector, $\lambda$ is projected out
by the orientifold projection.

The charged instanton spectrum is obtained from strings stretched
between the $E5$ instanton and the $D9$ background branes. The
corresponding cylinder amplitude is
\begin{align}
A_{E5-D9} &=  \frac{N p}{8}  \int_0^{\infty} \frac{dt}{t} \frac{\eta^2}{\vartheta_4^2} \eta^2 \sum_{\alpha \beta} c_{\alpha \beta}
\frac{\vartheta [{\alpha +1/2 \atop \beta }]^2}{\eta^2}
\frac{\eta}{\vartheta [{\alpha \atop \beta }]}
\{  P_1 P_2 P_3 \frac{\vartheta [{\alpha \atop \beta }]^3}{\eta^9}
+ \nonumber \\
& \left[ (-1)^{m_1} P_1 \frac{\vartheta [{\alpha \atop \beta }]\vartheta [{\alpha
    \atop \beta+1/2 }] \vartheta [{\alpha \atop \beta-1/2 }]}{\eta^3} \!+\!
 (-1)^{m_2} P_2 \frac{\vartheta [{\alpha \atop \beta-1/2 }]\vartheta [{\alpha
    \atop \beta }] \vartheta [{\alpha \atop \beta+1/2 }]}{\eta^3} \!+\! \right. \nonumber \\
&\left. +(-1)^{m_3} P_3 \frac{\vartheta [{\alpha \atop \beta +1/2}]\vartheta [{\alpha
    \atop \beta-1/2 }] \vartheta [{\alpha \atop \beta }]}{\eta^3} \right] \times \frac{1}{\eta^2} \frac{4 \eta^2}{\vartheta_2^2}  \} \ . \label{e5}
\end{align}
The massless states are described by the contributions
\begin{equation}
 A_{E5-D9}^{(0)} = N p \left( S_4 O_2 O_2 O_2 - O_4 C_2 C_2 C_2
 \right)
\ . \label{e7}
\end{equation}
In particular, the state $S_4 O_2 O_2 O_2$, coming from the NS sector, has a spinorial $SO(4)$ index
$\omega_{\alpha}$, whereas
$O_4 C_2 C_2 C_2$, coming from the R sector, is an $SO(4)$ scalar with a spinorial $SO(6)$ index or,
which is the same, a fundamental $SU(4)$ index $\mu^A$.
\subsection{E1 instantons}
The case of the $E1$ instantons is more subtle. Indeed, they wrap
one internal torus while they are orthogonal to the two remaining ones, thus feeling
the nontrivial effects of the freely-acting operations.  The explicit discussion can be limited
to the case of the $E1_1$ instantons, the other two
cases $E1_{2,3}$ being obviously completely similar.
It is useful to separately discuss the two distinct possibilities~:\\
i) the $E1_1$ instantons sit at one of the fixed points (tori) of the $g$
orbifold generator in the $y^1 \ldots y^6$ directions. \\
ii) the $E1_1$ instantons are located off the fixed points (tori) of the $g$ orbifold
generator in the $y^1 \ldots y^6$ directions.

It is worth to stress that, strictly speaking, the freely action $g$ has no fixed tori, due, of course, to the
shift along $T^1$.  However, since the instanton $E1_1$ wraps $T^1$,
while it is localized in the $(T^2,T^3)$ directions, it is convenient to analyze the
orbifold action in the space perpendicular to the instanton
world-volume.

In the following, we discuss the first configurations with the instantons on the fixed tori,
which are the relevant ones for matching the dual heterotic threshold corrections.
Since the freely-acting operations $(f,h)$ identify points in the internal space
perpendicular to the instanton world-volume, they enforce the presence of doublets of $E1_1$ instantons,
in complete analogy with similar phenomena happening in the case of background
D5 branes in \cite{partialbreak,aads2}.  Indeed, the $g$-operation is the only one acting in
a nontrivial way on the instantons. The doublet nature of the $E1_{1}$
instantons can be explicitly figured out in the following geometric way.
Let the location of the $E1_{1}$ instanton be fixed at a point of the $(y^3,y^4,y^5,y^6)$ space, which is left
invariant by the $g$-operation.  For instance, $| E1_1 \rangle = |0,0,\pi R_5 /2
,0 \rangle $. Then, the $f$
and $h$ operations both map the point $| E1_1 \rangle $ into
 $| E1'_1 \rangle = |\pi R_3,0, 3 \pi R_5 /2 ,0 \rangle$, so that an
orbifold invariant instanton state is provided by the combination (``doublet'')
\begin{equation}
\frac{1}{\sqrt 2} \  \left[ \ |0,0,\pi R_5 /2 ,0 \rangle \ + \
|\pi R_3,0, 3 \pi R_5 /2 ,0 \rangle \ \right] \ . \label{e07}
\end{equation}
The corresponding open strings can be stretched between fixed points and/or images, and can be
described by the following amplitudes
\begin{align}
A_{E1-E1} &=  \frac{q^2}{32} \int_0^{\infty} \frac{dt}{t} \frac{1}{\eta^2} \sum_{\alpha \beta} c_{\alpha \beta}
\frac{\vartheta [{\alpha \atop \beta }]}{\eta}
\{  P_1 (W_2 W_3 + W_2^{n+1/2} W_3^{n+1/2})
\frac{\vartheta [{\alpha \atop \beta }]^3}{\eta^9} \nonumber \\
& \ + \ (-1)^{m_1} P_1 \frac{\vartheta [{\alpha \atop \beta }]\vartheta [{\alpha
    \atop \beta+1/2 }] \vartheta [{\alpha \atop \beta-1/2 }]}{\eta^5} \frac{4 \eta^2}{\vartheta_2^2} \} \ , \label{e08}
\end{align}
\begin{align}
M_{E1-O9} &=  - \frac{q}{16} \int_0^{\infty} \frac{dt}{t} \frac{4 \eta^2}{\vartheta_2^2} \eta^2 \sum_{\alpha \beta}c_{\alpha \beta}
\frac{\vartheta [{\alpha \atop \beta+1/2 }] \vartheta [{\alpha \atop
      \beta-1/2 }]}{\eta^2} \frac{\eta}{\vartheta [{\alpha \atop
      \beta }]}  \{  (-1)^{m_1} P_1 W_2 W_3 \frac{\vartheta [{\alpha \atop \beta }]^3}{\eta^9} \ ,
\nonumber \\
& \ + \ P_1 \frac{\vartheta [{\alpha \atop \beta }]\vartheta [{\alpha
    \atop \beta+1/2 }] \vartheta [{\alpha \atop \beta-1/2 }]}{\eta^3} \frac{1}{\eta^2} \frac{4 \eta^2}{\vartheta_2^2}  \} \ .\label{e8}
\end{align}
Since only the $Z_2$ $g$-operation acts non-trivially on
the characters, it is convenient in this case to use covariant  $SO(4) \times SO(2) \times
SO(4)$ characters in order to describe the massless instanton
zero-modes. Due to the doublet nature of the instantons, particle
interpretation asks for a rescaling of the ``charge'' $q = 2 Q$, meaning that the tension of the elementary
instanton is twice the tension of the standard $D1$-brane.
The result is
\begin{equation}
A_{E1-E1}^{(0)} + M_{E1-O9}^{(0)} = \frac{Q (Q+1)}{2} \
(V_4O_2O_4 - C_4 C_2S_4) \ + \ \frac{Q (Q-1)}{2} \
( O_4 V_2 O_4 - S_4 S_2 S_4 ) \ . \label{e9}
\end{equation}
These zero-modes describe the positions $x^{\mu}$ of the $E1$ instantons in spacetime,
scalars $y^{i}$ along the torus wrapped by the instanton
and fermions $\Theta ^{{\dot \alpha},-,a }$,  $\Theta ^{{\alpha},+,a
}$.
The charged $E1_{1}$-$D9$ instanton spectrum is obtained from strings stretched
between the $E1$ instantons and the $D9$ background branes. The
corresponding cylinder amplitude is
\begin{align}
A_{E1-D9} &=  \frac{N q}{8}  \int_0^{\infty} \frac{dt}{t} \frac{\eta^2}{\vartheta_4^2} \eta^2 \sum_{\alpha \beta} c_{\alpha \beta}
\frac{\vartheta [{\alpha +1/2 \atop \beta }]^2}{\eta^2}
\frac{\eta}{\vartheta [{\alpha \atop \beta }]}
\{  P_1 \frac{ \vartheta [{\alpha \atop \beta }]
{\vartheta [{\alpha +1/2 \atop \beta }]}^2}{\eta^3}
\frac{\eta^2}{\vartheta_4^2} \nonumber \\
& + (-1)^{m_1} P_1 \frac{\vartheta [{\alpha \atop \beta }] \vartheta [{\alpha
+ 1/2  \atop \beta+1/2 }] \vartheta [{\alpha+1/2 \atop \beta-1/2 }]}{\eta^3} \frac{\eta^2}{\vartheta_3^2}  \} \label{e10}
\end{align}
The surviving massless states are now described by
\begin{equation}
 A_{E1-D9}^{(0)} = N Q \left( - O_4 S_2 O_4 \right)
\ , \label{e11}
\end{equation}
and correspond to the surviving ``would be'' world-sheet current algebra fermionic modes in the ``heterotic string''
interpretation (with $Q=1$ and $N=32$ \cite{dabh,witten2}).

The second configuration, where the $E1_1$ instantons are off the fixed points (tori) of the $g$ orbifold
generator in $y^1 \ldots y^6$, for instance $| E1_1 \rangle = |0,0,0,0 \rangle$, can be worked out as
well. In this case a quartet structure of instantons is present,  in a situation again similar to the ones
described in \cite{partialbreak,aads2}. Indeed,  $g$ produces the image $ g~:
\ |0,0,0,0 \rangle \rightarrow |0,0,\pi R_5,0 \rangle $, while $f$ and $h$ produce
two other images $ f~: \ |0,0,0,0 \rangle \rightarrow |\pi R_3,0,0,0
\rangle $ ,   $ h~: \ |0,0,0,0 \rangle \rightarrow |\pi R_3,0,\pi
R_5,0 \rangle $. In conclusion, the orbifold-invariant linear superposition of the
instanton images  is now the combination
\begin{equation}
\frac{1}{2} \left[ \ |0,0,0,0 \rangle + |0,0,\pi R_5,0 \rangle +   |\pi R_3,0,0,0
\rangle +  |\pi R_3,0,\pi R_5,0 \rangle \ \right] \ . \label{e12}
\end{equation}
For a given number of ``bulk'' $E1$ instantons, they have twice
the number of neutral (uncharged) fermionic zero modes as compared
to their ``fractional'' instantons cousins (\ref{e9}), whose
minimal number of uncharged zero modes is four.  On the other hand, their tension is twice
bigger.  If $n$ ``fractional'' $E1$ instanton doublets wrap the torus $T^i$, one expects a
contribution proportional to $e^{-4 \pi n T_i}$, whereas if they wrap half of the
internal torus, consistently with the shift identification, the contributions should be
proportional to $e^{-4 \pi (n+1/2) T_i}$. These considerations are perfectly in agreement
with the ${\cal N}=2$ nature of the threshold corrections
appearing in the heterotic computation (\ref{threhet}), (\ref{threhet4}) and (\ref{threhet6}).
On the other hand, the quartet structure of the  ``bulk'' instantons is probably incompatible
with them.  It should be also noticed that the absence of ${\cal
N}=1$ sectors contributing to the threshold corrections
(moduli-independent threshold corrections) on the heterotic side
reflects the fact that only the $f$ and $h$ action create instanton images.

A similar analysis to the one carried out in this section can be performed
for the more general type I $SO(n_o)\otimes SO(n_g)\otimes SO(n_f)\otimes SO(n_h)$ model
presented in section \ref{typeImodels}. However, we do not find any remarkable difference in nature between
different choices of $n_o$, $n_g$, $n_f$ and $n_h$, contrary to what the heterotic
dual model seems to suggest. It would be interesting to clarify this issue and to understand
why type I models differing only in the Chan-Paton charges lead to so different models in the
heterotic dual side.


\section{Fluxes and moduli stabilization}

\subsection{$\mathbb{Z}_2 \times\mathbb{Z}_2$ freely-acting orbifolds of twisted tori}

Background fluxes for the RR and NS-NS fields have been shown
to be relevant for lifting some of the flat directions of the
closed string moduli space. From the four dimensional effective field
theory perspective, the lift can be properly understood in terms
of a non-trivial superpotential encoding the topological
properties of the background. Many models based on ordinary
abelian orientifolds of string theory have
appeared in the literature (for recent reviews and references
see for instance \cite{grana}). Here we would like to extend this
construction to the case of orientifolds with a free action. The
motivation is two-fold. First, in these models the twisted sector
modes are massive, as has been previously shown. The same happens
for the open string moduli transforming in the adjoint. Second, we
have enough control over the non-perturbative regime, so that this model
provides us with a laboratory on which to explicitly test the
combined effect of fluxes and non-perturbative effects.

For the particular type I (heterotic) orbifolds considered here, the orientifold
projection kills a possible constant $H_3$ ($F_3$) background, so that the only possibilities left, apart from
non-geometric deformations, are RR (NSNS) 3-form fluxes and metric fluxes~\cite{kaloper,micu,new}. The latter correspond
to twists of the cohomology of the internal manifold $\mathcal{M}$,
\begin{equation}
d\omega_i=M_i{}^j\alpha_j+N^i{}_j\beta^j\ , \label{twist}
\end{equation}
where $\omega_i$ is a basis of harmonic 2-forms in $\mathcal{M}$,
and $(\alpha_i,\beta^j)$ a symplectic basis of harmonic 3-forms.
The resulting manifold $\tilde{\mathcal{M}}$ is in general no longer
Calabi-Yau, but rather it possesses $SU(3)$-structure~\cite{micu,minasian}. Duality arguments show, however, that the light
modes of the compactification in $\tilde{\mathcal{M}}$ can be suitably
described in terms of a compactification in $\mathcal{M}$, together with a
non-trivial superpotential $W_{twist}$ accounting for the
different moduli spaces.

Here we want to take a further step in the models of the previous sections and
to consider geometries which go beyond the toroidal one by adding metric fluxes to
the original torus. In terms of the global 1-forms of the torus,
the cohomology twist reads,
\begin{equation}
de^i=\frac{1}{2}f^i_{jk}e^j\wedge e^k\ , \label{twisttori}
\end{equation}
the resulting manifold being a group manifold $\tilde{\mathcal{M}}=G/\Gamma$ with
structure constants $f^i_{jk}$ and $\Gamma$ a discrete subgroup of $G$. Modding (\ref{twisttori})
by the orbifold action (\ref{gen1})-(\ref{gen3}) will in general put restrictions on the structure constants
$f^i_{jk}$ and the lattice $\Gamma$. More concretely, the surviving structure constants are

\begin{align*}
&\begin{pmatrix}f^2_{35}\\ f^4_{51}\\ f^6_{14}\end{pmatrix}=\begin{pmatrix}h_1\\ h_2\\ h_3\end{pmatrix}\ , \quad \begin{pmatrix}-f^1_{35}&f^4_{52}&f^6_{23}\\
f^2_{45}&-f^3_{51}&f^6_{14}\\
f^2_{36}&f^4_{61}&-f^5_{13}\end{pmatrix}=-\begin{pmatrix}b_{11}&b_{12}&b_{13}\\
b_{21}&b_{22}&b_{23}\\
b_{31}&b_{32}&b_{33}\end{pmatrix}\ , \\~\\
&\begin{pmatrix}f^1_{46}\\ f^3_{62}\\ f^5_{24}\end{pmatrix}=\begin{pmatrix}\bar h_1\\ \bar h_2\\ \bar h_3\end{pmatrix}\ , \quad \begin{pmatrix}-f^2_{46}&f^3_{61}&f^5_{14}\\
f^1_{36}&-f^4_{62}&f^5_{23}\\
f^1_{45}&f^3_{52}&-f^6_{24}\end{pmatrix}=-\begin{pmatrix}\bar b_{11}&\bar b_{12}&\bar b_{13}\\
\bar b_{21}&\bar b_{22}&\bar b_{23}\\
\bar b_{31}&\bar b_{32}&\bar b_{33}\end{pmatrix}\ ,
\end{align*}
as in an ordinary $\mathbb{Z}_2\times \mathbb{Z}_2$ orbifold. The Jacobi identity of the algebra
$G$ requires in addition $f^i_{[jk}f^m_{o]i}=0$~\cite{ss,kaloper}. The set of metric fluxes
transforms trivially under S-duality, so one can build heterotic-type I dual pairs by simply exchanging
$F_3\leftrightarrow H_3$.

The low energy physics of the $G/[\Omega_P \times \Gamma\times (\tilde{\mathbb{Z}}_2\times\tilde{\mathbb{Z}}_2)]$ compactification
can be then suitably described in terms of a
$T^6/[\Omega_P\times (\mathbb{Z}_2\times\mathbb{Z}_2)]$ compactification, with $\mathbb{Z}_2\times \mathbb{Z}_2$
being the freely-acting orbifold action described in section 2, together with a superpotential~\cite{hetflux},
\begin{equation}
W_{twist}=\sum_{i=1}^3T_i[-i\bar h_i+\sum_{j=1}^3\bar b_{ji}U_j+ib_{1i}U_2U_3+b_{2i}U_1U_3+ib_{3i}U_1U_2-h_iU_1U_2U_3]\ .
\end{equation}
Notice that the freely-acting $\tilde{\mathbb{Z}}_2\times\tilde{\mathbb{Z}}_2$ orbifold of the
full ten dimensional picture will in general differ from the freely-acting
$\mathbb{Z}_2\times \mathbb{Z}_2$ orbifold of the effective description. For illustration, consider the following simple example given by,
\begin{equation}
de^1=b_{11} e^3\wedge e^5\ , \quad de^2=de^3=de^4=de^5=de^6 = 0 \ .
\end{equation}
We may integrate these equations as,
\begin{equation}
e^1=dy^1+b_{11}y^3dy^5\ , \quad e^i=dy^i \quad \textrm{for }i\neq 1\ , \label{group1}
\end{equation}
so that $G$ is a fibration of $y^5$ over $y^1$. The lattice
$\Gamma$ is then suitably chosen as,
\begin{equation}
\Gamma: \quad \begin{cases}y^3\to y^3+1\ , \ \ y^1\to y^1-b_{11}y^5 & \\
y^i\to y^i+1 & \textrm{for }i\neq 3\end{cases}\ , \label{lattice}
\end{equation}
with $b_{11}\in \mathbb{Z}$ so that the vielbein vectors remain invariant under $\Gamma$ transformations.
Acting now with the orbifold generators (\ref{gen1})-(\ref{gen3}), it is not difficult to
convince oneself that in order the vielbein vectors to transform covariantly,
the orbifold generators have to be replaced by some new ones $\{\tilde g,\tilde f,\tilde h\}$ defined
as,
\begin{align}
&(y^1,\ y^2,\ y^3,\ y^4,\ y^5,\ y^6) \xrightarrow{\tilde g} (y^1+1/2,\ y^2,\ -y^3,\ -y^4,\ -y^5+1/2,\ -y^6) \ , \label{gen1p}\\
&(y^1,\ y^2,\ y^3,\ y^4,\ y^5,\ y^6) \xrightarrow{\tilde f} (-y^1+1/2+b_{11}y^5/2,\ -y^2,\ y^3+1/2,\ y^4,\ -y^5,\ -y^6) \ , \nonumber \\
&(y^1,\ y^2,\ y^3,\ y^4,\ y^5,\ y^6) \xrightarrow{\tilde h}
(-y^1-b_{11}y^5/2,\ -y^2,\ -y^3+1/2,\ -y^4,\ y^5+1/2,\ y^6) \ .
\nonumber
\end{align}

The generators $\{\tilde g,\tilde f,\tilde h\}$ still define a
$\mathbb{Z}_2\times \mathbb{Z}_2$ discrete group. Indeed,
requiring the quantization condition $b_{11}\in 2\mathbb{Z}$, one
can prove that $\tilde{g}^2=\tilde{h}^2=\tilde{f}^2=1$ and
$\tilde g\tilde f=\tilde f \tilde g=\tilde h$, $\tilde g\tilde
h=\tilde h\tilde g=\tilde f$, $\tilde h\tilde f=\tilde f\tilde
h=\tilde g$, up to discrete transformations of the lattice
$\Gamma$. Hence, the light modes of the $SU(3)$-structure
orientifold defined by the group manifold (\ref{group1}), together
with the lattice (\ref{lattice}) and the orbifold generators
(\ref{gen1p}), can be consistently described by a $T^6$
compactification with an orbifold action given by eqs.(\ref{gen1})
and a superpotential term,
\begin{equation}
W_{twist} \ = \ i b_{11} T_1U_2U_3 \ .
\end{equation}

\subsection{Moduli stabilization in a $S^3\times T^3/(\mathbb{Z}_2\times \mathbb{Z}_2)$
orbifold}

To illustrate the interplay between non-perturbative effects and
metric fluxes we consider in this section the following
one-parameter family of twists,
\begin{align*}
&de^1=\alpha e^4\wedge e^6\ , \quad  \quad de^2=\alpha e^4\wedge e^6 \ , \\
&de^3=\alpha e^6\wedge e^2\ , \quad  \quad de^4=\alpha e^6\wedge e^2 \ , \\
&de^5=\alpha e^2\wedge e^4\ , \quad  \quad de^6=\alpha e^2\wedge e^4 \ .
\end{align*}
The particular solution to these equations
\begin{align*}
e^1&=dy^1+e^2\ ,\quad & \quad e^2&=\sin(\alpha y^6)dy^4+\cos(\alpha y^6)\cos(\alpha y^4)dy^2\ , \\
e^3&=dy^3+e^4\ ,\quad & \quad e^4&=-\cos(\alpha y^6)dy^4+\sin(\alpha y^6)\cos(\alpha y^4)dy^2\ , \\
e^5&=dy^5+e^6\ ,\quad & \quad e^6&=dy^6+\sin(\alpha y^4)dy^2\ ,
\end{align*}
is corresponding to a product of a 3-sphere and a 3-torus.
Consistency requires $\alpha$ to be multiple of $2\pi$. On the
other hand, in this particular case the orbifold action remains
unaffected by the fluxes and is still given by
(\ref{gen1})-(\ref{gen3}).

We will also add a possible RR 3-form flux along the 3-sphere,
\begin{equation}
F_3=m\ e^2\wedge e^4\wedge e^6\ . \label{f3back}
\end{equation}
One may easily check that this flux, together with the above twists, does not
give rise to tadpole contributions.

The model can be effectively described by a $T^6/(\mathbb{Z}_2\times\mathbb{Z}_2)$
compactification with K\"ahler potential and superpotential,
\begin{align}
K&=-\log(S+S^*)-\sum_{i=1}^3\log(U_i+U_i^*)-\sum_{i=1}^3\log(T_i+T_i^*)\ ,\\
W&=m+\alpha \sum_{j=1}^3T_j(-i+U_j)+W_{np}(S,T_1,T_2,T_3,U_1,U_2,U_3)\ , \label{modelmin}
\end{align}
where we have introduced a generic non-perturbative superpotential
possibly depending on all moduli, as shown in the previous
sections\footnote{Perturbative corrections to the Kahler potential could also play a role in the moduli stabilization. We restrict here to the tree-level form of the Kahler potential, for the possible effect of $\alpha'$ or quantum
corrections to it,
see e.g. \cite{kahler}.}.

For $\textrm{Re }T_i \gg 1$ and $\textrm{Re }U_i \gg 1$ , the dependence of the non-perturbative
superpotential on the K\"ahler and complex structure moduli can be neglected,
$\partial_{U_i}W_{np}\simeq\partial_{T_i}W_{np}\simeq 0$, and the above superpotential
has a perturbative vacuum given by
\begin{align}
&\textrm{Im }U_i\simeq 1\ , & &\textrm{Re }W_{np}+m\simeq \alpha(\textrm{Re }T_i)(\textrm{Re }U_i) \ , \nonumber \\
&\textrm{Im }T_i\simeq 0\ , & &\textrm{Im }W_{np}\simeq 0 \ , \quad D_SW=0\ , \label{minimiza}
\end{align}
with $D_SW=\partial_SW-(S+S^*)^{-1}W$, as usual. Then, for $W_{np}$ the racetrack
superpotential (\ref{threrace3}), one may stabilize $S$ at a reasonably not too big coupling.

The model can be viewed in the S-dual heterotic side as an asymmetric $\mathbb{Z}_2\times \mathbb{Z}_2$
orbifold of some Freedman-Gibbons electrovac solution \cite{electro1,electro2}\footnote{We thank E.Kiritsis
for pointing out to us this connection.}. In particular, the full string ground state includes a
SU(2) Wess-Zumino-Witten model describing the radial stabilization of the 3-sphere by
$m$ units of $H_3$ flux, provided by $F_3\to H_3$ in (\ref{f3back}). In terms of the radii
$R_i$, $i=1\ldots 6$, equations (\ref{minimiza}) lead to
\begin{equation}
(R_2)^2=(R_4)^2=(R_6)^2\simeq \frac{\textrm{Re }W_{np}+m}{\alpha}\ ,
\end{equation}
whereas the radii of the 3-torus, $R_1$, $R_3$, $R_5$, remain as flat directions. Having
$\textrm{Re }T_i \gg 1$ and $\textrm{Re }U_i \gg 1$ then requires the volume of the 3-sphere to be much bigger
than the volume of the 3-torus, i.e. $m/\alpha\gg 1$.

\section*{Acknowledgments}
{We would like to thank C. Angelantonj, C. Bachas, M. Bianchi, E. Kiritsis, J.F.
Morales and A. Sagnotti for discussions. E.D. thanks CERN-TH and G.P. would like to thank  CPhT-Ecole
Polytechnique for the kind invitation and hospitality during the
completion of this work. G.P. would also like to thank P.
Anastasopoulos and F. Fucito for interesting
discussions. P.G.C. also thanks A. Font for discussions on related topics.
This work was also partially supported by INFN, by the INTAS contract 03-51-6346, by the
EU contracts MRTN-CT-2004-005104 and
MRTN-CT-2004-503369, by the CNRS PICS \#~2530, 3059 and 3747, by the MIUR-PRIN contract
2003-023852, by a European Union Excellence Grant,
MEXT-CT-2003-509661 and by the NATO grant PST.CLG.978785.}


\newpage

\appendix

\section{Normalization of string amplitudes}\label{normaliz}

For sake of brevity, throughout the paper we ignored the
overall factors coming from integrating over the noncompact
momenta. For arbitrary string tension $\alpha'$, the complete
string amplitudes ${\cal T}, {\cal K}, {\cal A}, {\cal M}$  are
related to the ones used in the main text by
\begin{eqnarray}
&& {\cal T} = \frac{1}{(4 \pi^2 \alpha')^2} \ T \quad , \quad
{\cal K} = \frac{1}{(8 \pi^2 \alpha')^2} \ K \ , \nonumber \\
&& {\cal A} = \frac{1}{(8 \pi^2 \alpha')^2} \ A \quad , \quad
{\cal M} = \frac{1}{(8 \pi^2 \alpha')^2} \ M \ . \label{norm1}
\end{eqnarray}
\section{Characters for $\mathbb{Z}_2 \times \mathbb{Z}_2$ orbifolds}
\label{apcharacters}

 In the light-cone RNS formalism, the vacuum amplitudes involve the following characters
\begin{align}
\tau_{oo}&=V_2I_2I_2I_2+I_2V_2V_2V_2-S_2S_2S_2S_2-C_2C_2C_2C_2 \ ,
\nonumber \\
\tau_{og}&=I_2V_2I_2I_2+V_2I_2V_2V_2-C_2C_2S_2S_2-S_2S_2C_2C_2 \ ,
\nonumber \\
\tau_{oh}&=I_2I_2I_2V_2+V_2V_2V_2I_2-C_2S_2S_2C_2-S_2C_2C_2S_2 \ ,
\nonumber \\
\tau_{of}&=I_2I_2V_2I_2+V_2V_2I_2V_2-C_2S_2C_2S_2-S_2C_2S_2C_2 \ ,
\nonumber \\
\tau_{go}&=V_2I_2S_2C_2+I_2V_2C_2S_2-S_2S_2V_2I_2-C_2C_2I_2V_2 \ ,
\nonumber \\
\tau_{gg}&=I_2V_2S_2C_2+V_2I_2C_2S_2-S_2S_2I_2V_2-C_2C_2V_2I_2 \ ,
\nonumber \\
\tau_{gh}&=I_2I_2S_2S_2+V_2V_2C_2C_2-C_2S_2V_2V_2-S_2C_2I_2I_2 \ ,
\nonumber \\
\tau_{gf}&=I_2I_2C_2C_2+V_2V_2S_2S_2-S_2C_2V_2V_2-C_2S_2I_2I_2 \ ,
\nonumber \\
\tau_{ho}&=V_2S_2C_2I_2+I_2C_2S_2V_2-C_2I_2V_2C_2-S_2V_2I_2S_2 \ ,
\nonumber \\
\tau_{hg}&=I_2C_2C_2I_2+V_2S_2S_2V_2-C_2I_2I_2S_2-S_2V_2V_2C_2 \ ,
\nonumber \\
\tau_{hh}&=I_2S_2C_2V_2+V_2C_2S_2I_2-S_2I_2V_2S_2-C_2V_2I_2C_2 \ ,
\nonumber \\
\tau_{hf}&=I_2S_2S_2I_2+V_2C_2C_2V_2-C_2V_2V_2S_2-S_2I_2I_2C_2 \ ,
\nonumber \\
\tau_{fo}&=V_2S_2I_2C_2+I_2C_2V_2S_2-S_2V_2S_2I_2-C_2I_2C_2V_2 \ ,
\nonumber \\
\tau_{fg}&=I_2C_2I_2C_2+V_2S_2V_2S_2-C_2I_2S_2I_2-S_2V_2C_2V_2 \ ,
\nonumber \\
\tau_{fh}&=I_2S_2I_2S_2+V_2C_2V_2C_2-C_2V_2S_2V_2-S_2I_2C_2I_2 \ ,
\nonumber \\
\tau_{ff}&=I_2S_2V_2C_2+V_2C_2I_2S_2-C_2V_2C_2I_2-S_2I_2S_2V_2 \ ,
\end{align}
where each term is a tensor product of the characters of the vector representation ($V_2$), the
scalar representation ($I_2$), the spinor representation ($S_2$) and the conjugate-spinor representation ($C_2$)
of the four $SO(2)$ factors that enter the light-cone restriction of the ten-dimensional Lorentz algebra.

\section{Details on the threshold correction computations}

\subsection{Threshold corrections in the type I \\ $SO(n_o) \otimes SO(n_g) \otimes SO(n_f)
    \otimes SO(n_h)$ models}
\label{appiso}

In order to implement the background field method, it is convenient to
express the orbifold characters in terms of the corresponding $\vartheta$-functions:
\begin{align}
\tau_{oo}+\tau_{og}+\tau_{oh}+\tau_{of}&=\frac{1}{2 \eta^4}(\vartheta_3^4-\vartheta_4^4-\vartheta_2^4-\vartheta_1^4)\ , \\
\tau_{oo}+\tau_{og}-\tau_{oh}-\tau_{of}&=\frac{1}{2 \eta^4}(\vartheta_2^2\vartheta_1^2+\vartheta_1^2\vartheta_2^2-\vartheta_4^2\vartheta_3^2+\vartheta_3^2\vartheta_4^2)\ , \\
\tau_{oo}-\tau_{og}+\tau_{oh}-\tau_{of}&=\frac{1}{2 \eta^4}(\vartheta_1\vartheta_2^2\vartheta_1+\vartheta_2\vartheta_1^2\vartheta_2+\vartheta_3\vartheta_4^2\vartheta_3-\vartheta_4\vartheta_3^2\vartheta_4)\ , \\
\tau_{oo}-\tau_{og}-\tau_{oh}+\tau_{of}&=\frac{1}{2 \eta^4}(\vartheta_2\vartheta_1\vartheta_2\vartheta_1+\vartheta_1\vartheta_2\vartheta_1\vartheta_2-\vartheta_4\vartheta_3\vartheta_4\vartheta_3+\vartheta_3\vartheta_4\vartheta_3\vartheta_4)\ .
\end{align}
Making use of the expansion (valid for even spin structure $\alpha)$)
\begin{equation}
\frac{\vartheta_{\alpha}(\epsilon\tau|\tau)}{\vartheta_1(\epsilon\tau|\tau)}=
\frac{1}{2\pi\epsilon\tau}\frac{\vartheta_{\alpha}}{\eta^3}+\frac{\epsilon\tau}
{4\pi}\frac{\vartheta_{\alpha}''}{\eta^3}+\ldots \ ,
\end{equation}
and the modular identities (\ref{modular1}) and (\ref{modular2}) in Appendix \ref{formulae}, the expansions
of the characters in terms of the (small) magnetic field or, equivalently, in terms of the $\epsilon$
of eq. (\ref{epsmode}), are
\begin{align}
&(\tau_{oo}+\tau_{og}+\tau_{oh}+\tau_{of}) (\epsilon \tau,\tau) \simeq
-\frac{i\epsilon\tau}{8
  \pi\eta^4}(\vartheta_3''\vartheta_3^3-\vartheta_4''\vartheta_4^3-\vartheta_2''\vartheta_2^3 )=0
\ , \nonumber \\
& (\tau_{oo}+\tau_{og}-\tau_{oh}-\tau_{of}) (\epsilon \tau,\tau)  =
(\tau_{oo}-\tau_{og}-\tau_{oh}+\tau_{of}) (\epsilon \tau,\tau) =
\nonumber \\
&\quad\quad  =(\tau_{oo}-\tau_{og}+\tau_{oh}-\tau_{of}) (\epsilon
\tau,\tau) \simeq - \frac{i\epsilon\tau}{8
  \pi\eta^4}(-\vartheta_4''\vartheta_4\vartheta_3^2+\vartheta_3''\vartheta_3\vartheta_4^2) =
\frac{i\pi\epsilon}{2} \tau\eta^2 \vartheta_2^2 \ . \ \label{del4}
\end{align}
The one-loop threshold corrections on any of the gauge group factors can therefore be written in the form
\begin{multline}
\Lambda_2=16\pi^2\int_0^{\infty}\frac{dt}{t}\{[2\textrm{Tr}(Q^2)-\textrm{Tr}(\gamma_g)
\textrm{Tr}(\gamma_g Q^2)](-1)^{m_1}P_1\ +\\
[2\textrm{Tr}(Q^2)-\textrm{Tr}(\gamma_f)\textrm{Tr}(\gamma_f
Q^2)](-1)^{m_2}P_2\ +
\ [2\textrm{Tr}(Q^2)-\textrm{Tr}(\gamma_h)\textrm{Tr}(\gamma_h Q^2)](-1)^{m_3}P_3\}\ ,
\end{multline}
where the action induced by the orbifold on the CP matrices, defined in (\ref{cp3}), has been used.
The last step is to compute the momentum sums $(-1)^mP$.  To this end, it is useful to reexpress (\ref{mp}) as
\begin{equation}
\Gamma \equiv \int_0^{\infty}\frac{dt}{t}(-1)^mP = \int_0^{\infty}\frac{dt}{t}\textrm{exp}\left[-\frac{\pi(\textrm{Re }T)}{4t(\textrm{Re }U)}\right]\sum_{m,m'}\textrm{exp}\left[-\pi(m-b)^TA(m-b)\right]\ ,
\end{equation}
with
\begin{equation}
m-b=\begin{pmatrix}m-\frac{i(\textrm{Re }T)}{2t(\textrm{Re }U)}\\
m'+\frac{i(\textrm{Re }T)(\textrm{Im }U)}{2t(\textrm{Re }U)}\end{pmatrix} \quad , \quad A=\frac{t}{(\textrm{Re }T)(\textrm{Re }U)}\begin{pmatrix}|U|^2 & \textrm{Im }U\\
\textrm{Im }U & 1\end{pmatrix}\ .
\end{equation}

Making use of the Poisson summation formula (\ref{poisson}) and
redefining $t\to 1/\ell$ in order to move to the transverse
channel picture, one gets
\begin{equation}
\Gamma = (\textrm{Re }T)\int_0^{\infty}d\ell\sum_{n_1,n_2}\textrm{exp}\left[-\frac{\pi\ell (\textrm{Re }T)}{\textrm{Re }U}[(n_1+\frac{1}{2}-n_2\textrm{Im }U)^2+(n_2\textrm{Re }U)^2]\right]\ . \label{integral}
\end{equation}
As expected, the integral contains infrared (IR) divergences as $\ell\to 0$,
corresponding to loops of massless modes.  It can be
regularized introducing an IR regulator $\mu$ via a factor
$F_{\mu}= (1-e^{- l / \mu^2 })$. Performing the integral in $\ell$ the result is
\begin{multline}
\Gamma = \lim_{\mu^2 \to 0} \left[\frac{\textrm{Re }U}{\pi}\sum_{n_1,n_2}\left(\frac{1}{(n_1+\frac{1}{2}-n_2\textrm{Im }U)^2+(n_2\textrm{Re }U)^2}-\right.\right. \\
\left.\left.-\frac{1}{(n_1+\frac{1}{2}-n_2\textrm{Im
      }U)^2+(n_2\textrm{Re }U)^2+\frac{\textrm{Re }U}{\pi \mu^2 \textrm{Re }T}}\right)\right]\ .
\end{multline}
Finally, using the Dixon, Kaplunovsky and Louis (DKL) formula~\cite{dkl} to evaluate
the sum over $n_1$, the expression become
\begin{equation}
\Gamma = -\sum_{n_2>0}\left[\frac{1}{n_2}\left(\frac{q^{n_2}-1}{q^{n_2}+1}+\frac{\bar{q}^{n_2}-1}{\bar{q}^{n_2}+1}\right)+\frac{2}{\sqrt{n_2^2+(
      (1/\pi (\textrm{Re
      }U)(\textrm{Re }T) \mu^2)}} \right]\ ,
\end{equation}
with $q\equiv \textrm{exp}[-2\pi U]$ and where we have taken $ \mu^2
\ll 1$ (in string units). A Taylor expansion (using eq. (\ref{series6})) produces
\begin{multline}
\Gamma = \sum_{n_2>0}\left(\frac{2}{n_2}-\frac{2}{\sqrt{n_2^2+(1/\pi(\textrm{Re
      }U)(\textrm{Re }T) \mu^2)}}\right)+2\sum_{n_2,m>0}\frac{(-1)^m}{n_2}q^{mn_2}+2\sum_{n_2,m>0}\frac{(-1)^m}{n_2}\bar{q}^{mn_2}\\
=\sum_{n_2>0}\left(\frac{2}{n_2}-\frac{2}{\sqrt{n_2^2+(1/\pi(\textrm{Re
      }U)(\textrm{Re }T) \mu^2)}}\right)-2\sum_{m>0}\textrm{log}(1-q^{2m})+2\sum_{m>0}\textrm{log}(1-q^{2m-1})\ + \ c.c.
\end{multline}
Taking the $\mu^2 \to 0$ limit and at the same time subtracting the
finite\footnote{The finite term can be actually reabsorbed into the value
of the gauge coupling at the compactification scale.} and the cut-off dependent parts,
in terms of the modular functions (\ref{series4}) and (\ref{series7}) one gets
\begin{equation}
\int_0^{\infty}\frac{dt}{t}(-1)^m F_{\mu} P = \textrm{log}
\left|\frac{\vartheta_4}{\eta^3}(2iU)\right|^2-\pi \textrm{Re
}U-\textrm{log}[(\textrm{Re }U)(\textrm{Re }T) \mu^2  ] \ .
\end{equation}
\subsection{Threshold corrections in the type I racetrack models}
\label{apprac}

The procedure for the racetrack models is completely analogous to the one in the previous section.
Plugging (\ref{del4}) into (\ref{raa}) and (\ref{ram}) one gets
\begin{multline}
\Lambda_{2,p}=16\pi^2\textrm{Tr}(Q^2)\int_0^{\infty}\frac{dt}{t}\\
\left[[(2-p)P_1-q(P_{m'+\frac{1}{2}}P_{m_1})](-1)^{m_1}+
(2-p)P_2(-1)^{m_2}+(2-p)P_3(-1)^{m_3}\right]\ ,
\end{multline}
where the $Q$ generator has been taken in the $SO(p)$ factor.
In this case there is a new lattice summation to compute, namely
\begin{multline}
\Gamma'= \int_0^{\infty}\frac{dt}{t}(-1)^{m}P_{m'+\frac{1}{2}}P_{m} = \\
=\int_0^{\infty}\frac{dt}{t}\sum_{m,m'}(-1)^m\textrm{exp}
\left[-\frac{\pi t}{(\textrm{Re }T)(\textrm{Re }U)}|m'+\frac{1}{2}-iUm|^2\right]=\\
=\int_0^{\infty}\frac{dt}{t}\textrm{exp}\left[-\frac{\pi (\textrm{Re }T)}{4t(\textrm{Re }U)}\right]\sum_{m,m'}\textrm{exp}[-\pi(m-b)^TA(m-b)]\ ,
\end{multline}
where now
\begin{equation}
m-b=\begin{pmatrix}m-\frac{i(\textrm{Re }T)}{2t(\textrm{Re }U)}\\
m'+\frac{i(\textrm{Re }T)(\textrm{Im U})}{2t(\textrm{Re }U)}+\frac{1}{2}\end{pmatrix}\quad , \quad
A=\frac{t}{(\textrm{Re }T)(\textrm{Re }U)}\begin{pmatrix}|U|^2& \textrm{Im }U\\
\textrm{Im }U& 1\end{pmatrix}\ .
\end{equation}
Thus, the integration in the transverse channel gives
\begin{equation}
\Gamma'= \frac{\textrm{Re }U}{\pi}\sum_{n_1,n_2}\frac{(-1)^{n_2}}{(n_1+\frac{1}{2}-n_2\textrm{Im }U)^2+(n_2\textrm{Re }U)^2}\ .
\end{equation}
Using again the (DKL) formula, after some algebra, the $\Gamma'$ can be written
\begin{equation}
\Gamma'= \sum_{n_2>0}\frac{1}{n_2}\left(\frac{q^{n_2}-1}{q^{n_2}+1}-\frac{q^{2n_2}-1}{q^{2n_2}+1}\right)\ + \ c.c. \ ,
\end{equation}
with $q=\textrm{exp}[-2\pi U]$. It should be noticed that in this case there is no need of an IR regulator for this sum.
In terms of modular functions the integral becomes
\begin{equation}
\int_0^{\infty}\frac{dt}{t}(-1)^{m}P_{m'+\frac{1}{2}}P_{m}
=\textrm{log}\left|\frac{\vartheta_4}{\eta^3}(4iU)\right|^2-\textrm{log}\left|\frac{\vartheta_4}{\eta^3}(2iU)\right|^2-\pi \textrm{Re }U\
\end{equation}
and the moduli dependent part of the gauge coupling threshold corrections is~
\begin{multline}
\Lambda_{2,p}=-16\pi^2\textrm{Tr}(Q^2)\left[(2-p)\sum_{j=1}^3\left(\pi
    \textrm{Re }U_j+\textrm{log}[(\textrm{Re }U_j)(\textrm{Re }T_j) \mu^2]-\textrm{log}\left|\frac{\vartheta_4}{\eta^3}(2iU_j)\right|^2\right)\right.\\
\left.+q\left(\textrm{log}\left|\frac{\vartheta_4}{\eta^3}(2iU_1)\right|^2-\textrm{log}\left|\frac{\vartheta_4}{\eta^3}(4iU_1)\right|^2+\pi \textrm{Re }U_1\right)\right]\ ,
\end{multline}
with a  $\beta$-function coefficient,
\begin{equation}
b_p \ = \ - \ 3 (p-2) \ ,
\end{equation}
that can be easily extracted from the previous expression.

\subsection{Threshold corrections in the heterotic models}
\label{apphet}

We consider separately the contributions from left- and right-mover oscillators
in (\ref{threskap}). The left-mover contributions read
\begin{multline}
\Lambda_{\textrm{left}}=\frac{1}{8\eta^3}\sum_{i=1}^3\left[\left(\partial_{\tau}\left(\frac{\vartheta_3}{\eta}\right)\vartheta_3\vartheta_4^2-\partial_{\tau}\left(\frac{\vartheta_4}{\eta}\right)\vartheta_4\vartheta_3^2\right)(-1)^{m_i+n_i}\Lambda_i\left|\frac{4\eta^2}{\vartheta_2^2}\right|^2+\right. \\
\left.+\left(\partial_{\tau}\left(\frac{\vartheta_3}{\eta}\right)\vartheta_3\vartheta_2^2-\partial_{\tau}\left(\frac{\vartheta_2}{\eta}\right)\vartheta_2\vartheta_3^2\right)\Lambda_i^{m_i+\frac{1}{2},n_i+\frac{1}{2}}\left|\frac{4\eta^2}{\vartheta_4^2}\right|^2-\right. \\
\left.-\left(\partial_{\tau}\left(\frac{\vartheta_2}{\eta}\right)\vartheta_2\vartheta_4^2-\partial_{\tau}\left(\frac{\vartheta_4}{\eta}\right)\vartheta_4\vartheta_2^2\right)(-1)^{m_i+n_i}\Lambda_i^{m_i+\frac{1}{2},n_i+\frac{1}{2}}\left|\frac{4\eta^2}{\vartheta_3^2}\right|^2\right]
\end{multline}
Making use of the identities (\ref{modular4}) - (\ref{eisenstein1}), we get after some small algebra
\begin{equation}
\Lambda_{\textrm{left}}=\frac{\pi i}{2\bar
\eta^6}\left[(-1)^{m_i+n_i}\hat
Z_i\overline{\vartheta}_3^2\overline{\vartheta}_4^2-\hat
Z_i^{m_i+\frac{1}{2},n_i+\frac{1}{2}}\overline{\vartheta}_2^2\overline{\vartheta}_3^2-(-1)^{m_i+n_i}\hat
Z_i^{m_i+\frac{1}{2},n_i+\frac{1}{2}}\overline{\vartheta}_2^2\overline{\vartheta}_4^2\right]\
,
\end{equation}
where the toroidal lattice sums $\hat Z_i\equiv |\eta|^4\Lambda_i$
are provided by (\ref{lat11}) - (\ref{lat13}), after Poisson
resummation in $m_1$ and $m_2$.

Regarding the contributions from the right-mover fermionic
oscillators, we get
\begin{equation}
\Lambda_{\textrm{right}}=\left(Q^2_{SO(32)}-\frac{1}{4\pi\tau_2}\right)\frac{1}{2}\sum_{a,b}\frac{\overline{\vartheta}\left[{a
\atop
b}\right]^{16}}{\bar{\eta}^{16}}=-\frac{1}{8\pi^2}\frac{\overline{\vartheta}\left[{a
\atop b}\right]''\overline{\vartheta}\left[{a \atop
b}\right]^{15}}{\bar{\eta}^{16}}-\frac{1}{8\pi\tau_2}\sum_{a,b}\frac{\overline{\vartheta}\left[{a
\atop b}\right]^{16}}{\bar{\eta}^{16}}\ .
\end{equation}
Making use of relations (\ref{modular5}) - (\ref{eisenstein2}), these
terms can be rearranged in the very compact expression
\begin{equation}
\Lambda_{\textrm{right}}=\frac{\overline{E}_4(\overline{E}_4\overline{\hat
E}_2-\overline{E}_6)}{12\bar \eta^{16}}\ ,
\end{equation}
corresponding to the modular covariant derivative of $\overline{E}_8$.

Putting all together we then arrive to the final expression for the
gauge kinetic threshold corrections to the SO(32) heterotic model,
\begin{multline}
\Lambda_2=\frac{i}{4\pi}\int_{\mathcal{F}}\frac{d^2\tau}{\tau_2}\Lambda_{\textrm{left}}\Lambda_{\textrm{right}}=-\frac{1}{96}\int_{\mathcal{F}}\frac{d^2\tau}{\tau_2}\sum_{i=1}^3\left[
(-1)^{m_i+n_i}\hat
Z_i\overline{\vartheta}_3^2\overline{\vartheta}_4^2-
\hat Z_i^{m_i+\frac{1}{2},n_i+\frac{1}{2}}\overline{\vartheta}_2^2\overline{\vartheta}_3^2-\right. \\
\left.-(-1)^{m_i+n_i}\hat
Z_i^{m_i+\frac{1}{2},n_i+\frac{1}{2}}\overline{\vartheta}_2^2\overline{\vartheta}_4^2\right]\frac{\overline{E}_4(\overline{\hat
E}_2\overline{E}_4-\overline{E}_6)}{\bar{\eta}^{24}}\ .
\end{multline}
In the limit of large volume, $\textrm{Re }T_i\gg 1$, or equivalently $q\to 0$
and $n_i=0$, only degenerate orbits consisting of $A$ matrices (\ref{lat13})
with zero determinant in the sector $(h,g)=(1,0)$ contribute to the toroidal
lattice sums. Following \cite{dkl}, then we can pick an element
$A_0$ in each orbit and to integrate its contribution over the
image under $V$ of the fundamental domain, for all $V\in SL(2)$
yielding $A_0V\neq A_0$. The representatives can be chosen to be,
\begin{equation}
A_0=\begin{pmatrix}0 & j+\frac{1}{2}\\ 0 & p\end{pmatrix}\ ,
\end{equation}
enforcing the identification
\begin{equation}
(j, p) \sim (-j-1,-p)\ .
\end{equation}
With this representation, $A_0V'=A_0V''$ if and only if
$$V'=\begin{pmatrix}1& m\\ 0& 1\end{pmatrix}V'' \ . $$
Therefore, the contributions are integrated over
$\{\tau_2>0,|\tau_1|<\frac{1}{2}\}$, and the double covering is
taking into account by summing over all $p$ and $j$,
\begin{equation}
I_d=(\textrm{Re
}T)\int_{-\frac{1}{2}}^{\frac{1}{2}}d\tau_1\int_0^{\infty}\frac{d\tau_2}{\tau_2^2}\sum_{j,p}\textrm{exp}\left(-\frac{\pi
\textrm{Re }T}{\tau_2\textrm{Re }U}|j+\frac{1}{2}+iU_ip|^2\right)\
.
\end{equation}
This is exactly the same expression as (\ref{integral}), so the contributions
of the degenerate orbits perfectly match the perturbative type I threshold corrections,
\begin{equation}
I_d=\textrm{log}\left|\frac{\vartheta_4}{\eta^3}(2iU_i)\right|^2-\pi
\textrm{Re }U_i-\textrm{log}[(\textrm{Re }U_i)(\textrm{Re }T_i)]\
.
\end{equation}
Analogously, in the limit $q\to 0$ but $n_i\neq 0$ also the
non-degenerate orbits in the sector $(h,g)=(1,0)$ contribute. The representative in this class can be
chosen to have the form
\begin{equation}
A_0=\begin{pmatrix}k & j+\frac{1}{2}\\ 0 & p\end{pmatrix}\ ,
\end{equation}
with $k>j\geq0, \ p\neq 0$. For these, $V'\neq V''$ implies
$A_0V'\neq A_0V''$, and therefore these contributions must be
integrated over the double cover of the upper half plane
($\tau_2>0$),
\begin{equation}
I_{nd}=2(\textrm{Re }T)\sum_{0\leq j < k, p\neq 0}e^{2\pi
Tkp}\int_{-\infty}^{\infty}d\tau_1\int_0^{\infty}\frac{d\tau_2}{\tau_2^2}
(-1)^k \textrm{exp}\left[-\frac{\pi\textrm{Re
}T}{\tau_2\textrm{Re }U}|k\tau+j+\frac{1}{2}+ipU|^2\right]\ .
\end{equation}
Evaluating the gaussian integral over $\tau_1$ and summing on $j$,
one gets
\begin{equation}
I_{nd} = 2\sum_{0<k,p\neq 0}e^{2\pi
Tkp}\int_0^{\infty}d\tau_2\sqrt{\frac{(\textrm{Re }U)(\textrm{Re
}T)}{\tau_2^3}} (-1)^k \textrm{exp}\left[-\frac{\pi \textrm{Re
}T}{\tau_2\textrm{Re }U}(k\tau_2+p\textrm{Re }U)^2\right]\ ,
\end{equation}
and the contribution of this sector becomes
\begin{equation}
I_{nd}= \textrm{log}|\frac{\vartheta_4}{\eta^3} (2iT)|^2 - \pi
\textrm{Re }T\ . \label{ind}
\end{equation}
It corresponds to $E1$ instanton corrections in the type
I $SO(32)$ dual model. Indeed, expanding the $\eta$-function in
(\ref{ind}), $I_{nd}$ can be expressed as
\begin{equation}
I_{nd}=-2 \sum_{n=1}^{\infty} (-1)^n \textrm{log }(1-e^{-2\pi n T})\ + \
c.c.\ ,
\end{equation}
which should correspond to a sum over the contributions of
$E1$-instantons wrapping $n$ times the (1,1)-cycle associated to
$T$, a fact that would be very interesting to verify explicitly.
Notice that the dependence on $T$ perfectly agrees with the
general arguments in \cite{germans} for the mirror type IIA
picture.


\section{Some useful formulae}\label{formulae}

{- Poisson summation formula:}
\begin{equation}
\sum \textrm{Exp}\left[-\pi(m-b)^TA(m-b)\right]=\frac{1}{\sqrt{\textrm{det }A}}\sum\textrm{Exp}\left[-\pi n^TA^{-1}n+2i\pi b^Tn\right] \ . \label{poisson}
\end{equation}

\noindent {- Modular identities:}
\begin{align}
&\vartheta_3''\vartheta_3^3-\vartheta_4''\vartheta_4^3-\vartheta_2''\vartheta_2^3
 \ = \ 0 \ , \label{modular1}\\
&\vartheta_3''\vartheta_3\vartheta_4^2-\vartheta_4''\vartheta_4\vartheta_3^2=-4\pi^2\eta^6\vartheta_2^2 \ , \label{modular2}\\
&\vartheta_2\vartheta_3\vartheta_4=2\eta^3\ , \label{modular4}\\
&\vartheta_2''=4\pi i \partial_\tau \vartheta_2=-\frac{\pi^2}{3}\vartheta_2(E_2+\vartheta_3^4+\vartheta_4^4)\ , \label{modular5} \\
&\vartheta_3''=4\pi i \partial_\tau \vartheta_3=-\frac{\pi^2}{3}\vartheta_3(E_2+\vartheta_2^4-\vartheta_4^4)\ , \\
&\vartheta_4''=4\pi i\partial_\tau \vartheta_4=-\frac{\pi^2}{3}\vartheta_4(E_2-\vartheta_2^4-\vartheta_3^4)\ , \label{modular6}
\end{align}

\noindent {- Eisenstein series:}
\begin{align}
&E_2=\hat E_2+\frac{3}{\pi\tau_2}=\frac{12}{i\pi}\partial_\tau\textrm{log }\eta = 1-24q-\ldots \ , \label{eisenstein1} \\
&E_4=\frac{1}{2}(\vartheta_2^8+\vartheta_3^8+\vartheta_4^8) = 1+240q+\ldots \ , \\
&E_6=\frac{1}{2}(\vartheta_2^4+\vartheta_3^4)(\vartheta_3^4+\vartheta_4^4)(\vartheta_4^4-\vartheta_2^4)=1-540q-\ldots \ ,\\
&E_8=E_4^2=\frac{1}{2}(\vartheta_2^{16}+\vartheta_4^{16}+\vartheta_3^{16})=1+480q+\ldots \ , \\
&E_{10}=E_4E_6=-\frac{1}{2}\left[\vartheta_2^{16}(\vartheta_3^4+\vartheta_4^4)+\vartheta_3^{16}(\vartheta_2^4-\vartheta_4^4)-\vartheta_4^{16}(\vartheta_2^4+\vartheta_3^4)\right]\
,\label{eisenstein2}
\end{align}

\noindent {- Series expansions:}
\begin{align}
&\textrm{log}(1-Q)=-\sum_{n=1}\frac{Q^n}{n} \ , \label{series1} \\
&\textrm{log}(1+Q)=-\sum_{n=1}(-1)^n\frac{Q^n}{n} \ , \\
&\textrm{log }\vartheta_2=\textrm{log }2q^{1/8}+\sum_{n=1}\textrm{log}(1-q^n)+2\sum_{n=1}\textrm{log}(1+q^n) \ , \\
&\textrm{log }\vartheta_4=\sum_{n=1}\textrm{log}(1-q^n)+2\sum_{n=1}\textrm{log}(1-q^{n-\frac{1}{2}}) \ , \label{series7} \\
&\textrm{log }\eta = \textrm{log }q^{1/24}+\sum_{n=1}\textrm{log}(1-q^n) \ , \label{series4} \\
&\frac{1+Q}{1-Q}=1+\sum_{m=1}Q^m \ , \label{series5}\\
&\frac{Q-1}{Q+1}=-1-2\sum_{m=1}(-1)^m Q^m \ . \label{series6}
\end{align}


\end{document}